# Strengthening and interface-mediated plastic co-deformation in an ultrafine Cr–Ni eutectic: A nanomechanical investigation

Arkajit Ghosh [a,#], Mustafa Tobah [a], Jianing Zhou [b], Jian Wang [b], Amit Misra [a,#]

[a] Department of Materials Science and Engineering, University of Michigan – Ann Arbor, MI 48109, USA

[b] Department of Mechanical and Materials Engineering, University of Nebraska – Lincoln, NE 68588, USA

[#] Corresponding author: arkajitg@umich.edu, amitmis@umich.edu

## *Abstract*

Ultrafine eutectic heterostructures provide a stringent test of plasticity in high-strength materials, where deformation must be accommodated through interfaces and strain gradients. Room temperature ductility is typically limited by premature fracture of the hard phase, leaving open the fundamental questions regarding the interface spacing, atomic structure and local chemistry that enable plastic co-deformation. Here we address this question using a model system of Cr–Ni binary alloy, processed via electron-beam powder bed fusion that produces a lamellar eutectic microstructure of Cr-rich body-centered cubic (BCC) and Ni-rich face-centered cubic (FCC) phases, with an average interlamellar spacing of ~450 nm. Atomic-resolution scanning transmission electron microscopy (STEM) revealed a stepped semi-coherent FCC/BCC interface with the Kurdjumov–Sachs (K–S) orientation relationship and Ni enrichment confined to a few atomic planes on the BCC side. *In situ* micro-scale compression and tension tests in SEM demonstrate high flow stresses coupled with large plastic strains without cracking, indicating stable accommodation of plastic incompatibility. Correlative TEM/HR-STEM establishes a deformation sequence: initial plasticity is dominated by strain-gradient driven dislocation accumulation in the FCC lamellae adjacent to interfaces, followed by deformation twinning in FCC Ni and local interfacial shear and reorientation. The BCC phase subsequently develops a high density of mobile dislocations. Atomistic modeling has been employed to understand the influence of the FCC/BCC interface atomic structure and chemistry on the slip activation in the hard phase. These findings show that nanoscale confinement, stepped K–S interfacial structure, and interfacial chemistry collectively promote dislocation glide in a hard phase below its monolithic brittle to ductile transition temperature, and plastic co-deformation at high flow strengths.

## 1. Introduction

Ultrafine-scale eutectics are model systems for fundamental understanding of deformation mechanisms in interface-dominated microstructures with disparate phases, also referred to as heterostructured materials [1, 2, 3]. Plastic flow in heterostructures containing hard phases becomes inherently non-uniform: elastic–plastic mismatch between adjoining phases imposes plastic strain gradients and necessitates the storage of geometrically necessary dislocations (GNDs) to maintain compatibility [4]. Ashby's seminal analysis of plastically non-homogeneous materials established that such gradients could raise hardening far beyond that of homogeneous crystals by requiring excess dislocation content to accommodate incompatibility [5]. This concept forms the mechanistic basis of modern strain-gradient plasticity and size-effect frameworks, which link strengthening to GND density through Taylor-type relations and explicitly introduce length scales that become important at micrometer and smaller dimensions [6, 7]. However, heterostructured materials containing hard phases are frequently limited by a fundamental failure mode: the hard phase (often BCC, diamond cubic, or intermetallic) fractures before it can sustain plastic flow, thereby preventing plastic co-deformation [8, 9, 10]. In BCC metals, this limitation is tightly connected to the ductile-to-brittle transition, where cleavage competes with dislocation-mediated plasticity [11] and dislocation mobility, particularly of screw segments, can be temperature-limiting [12]. A recent work has demonstrated that in pure BCC metals, such as Cr, the relative mobility of screw versus edge dislocations can control the transition by determining whether dislocation sources can operate efficiently and sustain multiplication [13]. These insights motivate a central scientific question for fine-scale hetero-structures: what microstructural and interfacial conditions enable the hard phase to activate slip and co-deform, rather than crack, under ambient loading?

In layered metallic systems, when the interface spacing is reduced to nanometer-scale, deformation is governed by interface-controlled unit processes – including confined layer slip, dislocation transmission/absorption, and interface-mediated nucleation, rather than bulk-like forest hardening alone [14, 15, 16]. In eutectic systems, classical growth theory provides a basis to tune lamellar spacing through coupled growth kinetics, offering a pathway to systematically vary the governing length scale for compatibility and GND storage [17]. For FCC/BCC couples, the interface crystallography adds an additional degree of control: orientation relationships (ORs) that yield relatively lower energy interphase boundaries that are often faceted/stepped, where steps can be interpreted within a disconnection framework (interfacial line defects with both dislocation and step character) that governs interfacial shear accommodation and defect transfer [18, 19]. Processing routes that refine spacing while preserving controlled interface character are therefore essential for isolating mechanism.

Electron-beam powder bed fusion (EPBF) is particularly attractive because it combines repeated rapid solidification cycles with elevated build temperatures under vacuum, enabling ultrafine microstructures due to high undercooling while potentially promoting interfacial ordering and chemical redistribution during thermal cycling [20, 21]. In this work, EPBF is used to produce fine-scale Cr-rich BCC / Ni-rich FCC lamellar eutectic microstructure. By combining *in situ* nanomechanical testing with correlative transmission electron microscopy (TEM) and high resolution scanning transmission electron microscopy (HR-STEM), and atomistic modeling, we interrogate the transition from soft-phase dominated flow to true plastic co-deformation, with particular emphasis on unit mechanisms such as strain hardening and deformation twinning in the FCC phase, and interface-enabled activation of mobile dislocations in the BCC that collectively govern the onset of cooperative plasticity at high strength levels.

## 2. *Materials and Methods*

### 2.1. *Specimen preparation*

Elemental Cr and Ni powders were blended at a weight ratio of 57:43. A slightly hypereutectic (Cr-rich) composition was selected to promote a fully eutectic microstructure by compensating for the kinetic shift of the equilibrium eutectic composition (50:50) toward the hypereutectic side; this shift occurs under rapid solidification when the liquidus and solidus curves approach the $T_0$ (temperature at which liquid and solid phases of the same composition have equal Gibbs free energy due to complete solute trapping conditions) lines [22]. Electron beam powder bed fusion (EPBF) was performed in a Freemelt 3D printer using optimized parameters: beam spot size of 250 µm, scanning speed of 200 m/s, beam power of 400 W, hatch spacing of 500 µm, and layer thickness of 50 µm. Prior to printing, the build plate was preheated to ~900 °C by electron-beam heating rather than by an external substrate heater; this temperature was not actively maintained throughout the build. Based on a thermocouple in contact with the bottom of the build plate, the plate temperature measured decreased from ~900 °C to ~700 °C during the ~3-hour build required to fabricate nine samples with a height of 7.5 mm. After completion of the build, the samples were allowed to cool slowly under vacuum, reaching to room temperature over approximately one day. Tiny blocks of samples measuring 10 mm x 10 mm x 7.5 mm (length x width x height) were printed. Pure Cr and Ni reference samples were prepared by arc-melting by elemental (99.995% pure) Cr and Ni pieces in inert atmosphere in an AMAZEMET arc-melter on water cooled Cu hearth.

### 2.2. *Microstructure characterization*

Scanning electron microscopy (SEM) was performed using a TFS Helios 650 NanoLab equipped with backscattered electron (BSE) detectors, operated at an accelerating voltage of 5 kV and a beam current of 0.2 nA. Cross-sectional TEM foils were prepared by focused ion beam (FIB) lift-out and subsequently thinned to a final thickness of ~50 nm. Atomic- to near-atomic-resolution scanning transmission electron microscopy (STEM) was conducted on an aberration-corrected TFS Spectra (S)TEM operated at 300 kV with a 0.05 nm probe size and a screen current of 75 pA. High-angle annular dark-field (HAADF), annular dark-field (ADF), and bright-field (BF) detectors were employed for detailed microstructural characterization. Two-beam-condition bright-field and dark-field imaging was performed using a TEM operated at 200 kV. Elemental mapping was carried out by energy-dispersive X-ray spectroscopy (EDX) using a DualX detector with the screen current increased to 100 pA. TemCompanion: An open-source multi-platform GUI program was employed for qualitative strain measurement using geometric phase analysis (GPA) [23].

### 2.3. Nanomechanical testing

*In situ* dog-bone tensile and micropillar compression tests were performed in a TESCAN MIRA 3 SEM using a Hysitron PI 89 Picoindenter. Dog-bone tensile specimens (10 μm gauge length × 3 μm gauge width × 3 μm gauge thickness) and micropillars (9 μm diameter × 22 μm height) were fabricated by Xe FIB in a TFS Hydra 5 PFIB-SEM using standard milling patterns developed earlier for optimal geometry to minimize stress concentrations [24]. Both tensile and compression tests were conducted under displacement control at a nominal strain rate of 0.5% /s, with a maximum applied nominal strain of 35%. Tensile tests were conducted using a diamond gripper with a ~10 μm jaw opening, prepared by FIB milling of a Berkovich indenter. For micropillar compression, a Bruker 20 μm flat-punch, 60° conical diamond probe installed on a high-load transducer was used. Nanoindentation experiments were performed in load-controlled mode with a nominal peak load of 50 mN and loading/hold/unloading times of 5/2/5 s, respectively. All the tests were repeated three times to ensure reproducibility of the results.

### 2.4. Atomistic modeling

Density functional theory (DFT) calculations were performed using the Vienna *Ab initio* Simulation Package (VASP) [25] with plane wave basis set. The generalized gradient approximation (GGA) with the Perdew-Burke–Ernzerhof (PBE) parametrization [26] was used for the electronic exchange and correlation interaction. A Γ-centered Monkhorst-Pack scheme was adopted in all the slab models and specific k-point mesh sizes were selected based on the sizes of the systems. For the Kohn-Sham ground state calculations, a cutoff energy of 400 eV for geometry optimization and minimization purpose were considered based on the Projector Augmented Wave (PAW) pseudopotentials. Spin polarized calculations were performed for all systems containing Cr and Ni atoms. The total energy convergence thresholds for the self-consistent iterations were set to $10^{-5}$ eV and $10^{-6}$ eV for the structural relaxation and the final energy evaluation steps, respectively. The geometry optimizations were performed until the residual forces acting on each atom were relaxed to less than $10^{-2}$ eV/Å. The electronic partial occupancies were determined using the Methfessel-Paxton scheme for structural relaxations, while the tetrahedron method with Blöchl corrections was utilized for the final energy evaluations.

## 3. Results

### 3.1. Ultrafine eutectic FCC/BCC heterostructure

EPBF processing of blended Cr and Ni powders produced an ultrafine lamellar eutectic microstructure within the outlined eutectic colonies shown in Figure 1a, with an average interlamellar spacing of $436 \pm 56$ nm. This refinement is consistent with the high solidification rates and associated kinetic undercooling of EPBF relative to near-equilibrium eutectic growth. In contrast, Kossowsky *et al*. reported an average spacing of ~10 μm for directionally solidified Ni–Cr eutectic, reflecting the much lower undercooling and growth rates in unidirectional solidification [27]. A higher-magnification STEM-HAADF image (Figure 1b1) resolves the lamellae within a eutectic colony, while simultaneous STEM-EDX maps (Figure 1b2–1b4) delineate the two phases. The area profile across the interfaces (Figure 1c, area of interest in Figure 1b1) confirms that the brighter HAADF lamellae are Ni-rich containing a substantial Cr solute fraction, whereas the darker lamellae are Cr-rich with only minor Ni. This partitioning is consistent with the Cr–Ni phase diagram (Figure 1d), which indicates high Cr solubility in Ni (35 wt.% at 900 °C, build plate temperature) but limited Ni solubility in Cr (~5 wt.%) at that temperature. Because the lamellae are sub-micron, CBED was used to establish the zone-axis patterns locally prior to global SAED analysis; the resulting patterns index to BCC $[\bar{1}\ 1\ 1]$ for the Cr-rich phase and FCC $[0\ 1\ 1]$ for the Ni-rich (Figure 1e). Quantitative phase composition (EDX), crystallography, characteristic length scales, and volume fractions are summarized in Table 1.

Table 1: Characterization of the phases in ultrafine eutectic processed by EPBF

| Phase | Composition (wt.%) | Crystal structure | Phase width (nm) | Volume fraction |
|---|---|---|---|---|
| Ni-rich | Ni: $64 \pm 2$; Cr: $36 \pm 2$ | FCC | $320 \pm 41$ | ~0.70 |
| Cr-rich | Cr: $95 \pm 1$; Ni: $5 \pm 1$ | BCC | $114 \pm 33$ | ~0.30 |

SAED pattern collected from a eutectic colony with several lamellae as displayed in Figure 1f reveals a well-defined Kurdjumov–Sachs (K–S) OR given as:

$$(1\ \bar{1}\ 1)_{FCC} \parallel (1\ 1\ 0)_{BCC};\ [0\ 1\ 1]_{FCC} \parallel [\bar{1}\ 1\ 1]_{BCC}$$

HR-STEM HAADF micrograph in Figure 1g captures the interface between FCC Ni-rich phase and BCC Cr-rich phase at atomic level. The interface is periodically stepped in nature, and the interface plane makes an effective angle of 15° with respect to the parallel planes (i.e., common plane). The stepped FCC/BCC interface is consistent with faceted, semi-coherent interphase boundary formation during coupled eutectic growth, where the interface adopts low-energy terrace planes that satisfy the K–S OR and advances by lateral ledge (step) propagation rather than as an atomically rough front. Each step effectively acts as an interfacial disconnection, allowing the boundary to maintain near-coherency over terraces while concentrating misfit accommodation into localized step/misfit-dislocation regions, thereby reducing the overall interfacial energy [28]. Although there are no prior studies on EB-PBF Cr-Ni eutectic, the stepped K–S interface observed here is broadly consistent with reported FCC/BCC interfaces in cast/heat treated Ni–Cr alloys. Luo and Weatherly [29, 30] showed that Cr-rich BCC precipitates in Ni–45 wt.% Cr adopt a K–S OR with the FCC matrix and preferentially form low-energy $\{121\}_{FCC}$-type habit planes. Atomistic simulations by Chen, Farkas, and Reynolds [31] further showed that the corresponding FCC/BCC Ni–Cr interface has a relatively low interfacial energy and consists of structural ledges/facets at the atomic scale. In the present EB-PBF printed eutectic, the SAED pattern and HR-STEM image similarly reveal a K–S-type orientation relationship and an atomically-stepped FCC/BCC interface. However, because the present interface forms during rapid solidification followed by repeated thermal cycling and slow cooling, the exact local habit-plane inclination and step geometry need not be identical to those reported for solid-state precipitation in well-annealed Ni–Cr alloys.

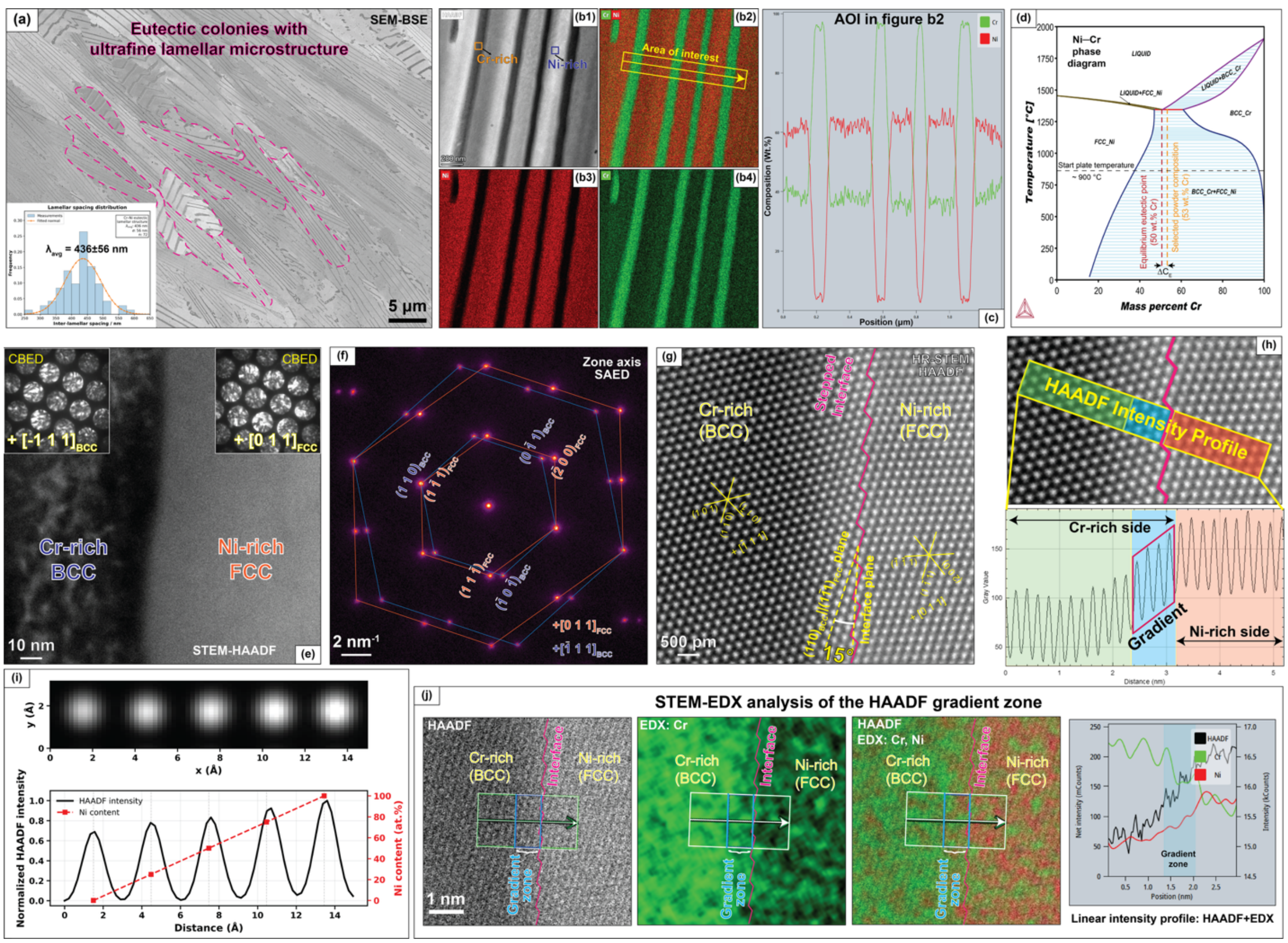

Figure 1: (a) Representative SEM-BSE image of the EPBF-processed Cr–Ni eutectic colonies with ultrafine lamellae, having average interlamellar spacing of $436 \pm 56$ nm. The inset shows the corresponding lamellar spacing distribution. (b1–b4) STEM HAADF image and corresponding STEM-EDX maps identifying Ni-rich and Cr-rich lamellae. (c) STEM-EDX line profile acquired along the line indicated in (b1), confirming Ni-rich lamellae containing substantial Cr in solid solution and Cr-rich lamellae with minor Ni. (d) Cr–Ni phase diagram highlighting asymmetric mutual solubility at 900 °C (EPBF build-plate temperature), with high Cr solubility in Ni and limited Ni solubility in Cr. (e) CBED indexing of FCC [0 1 1] (Ni-rich) and BCC $[\bar{1}\ 1\ 1]$ (Cr-rich) lamellae. (f) SAED showing the K–S OR. (g) HR-STEM HAADF revealing a periodically stepped semi-coherent interface (~15° inclination). (h) HAADF-intensity profile indicating a ~ 0.5-1 nm near-interface Z-contrast gradient on the BCC side, suggesting Ni enrichment. (i) abTEM-simulated STEM-HAADF image and intensity profile for a simplified single-row Cr–Ni interface model with progressively increasing Ni content from left to right. The simulated intensity increases with Ni fraction, corroborating the experimentally observed HAADF gradient. (j) STEM-EDX maps and corresponding line profile from the interfacial gradient zone, showing Ni enrichment on the Cr-rich side of the FCC/BCC interface.

In EPBF, electron-beam preheating raises the build plate to a high initial temperature (~900 °C), and the build remains hot (~700 °C) during printing despite gradual cooling of the plate. In combination with repeated thermal cycling and slow post-build cooling under vacuum, this thermal history can promote interfacial reconstruction and defect ordering, such as step bunching and a more regular misfit network, compared with a rapidly quenched state [32, 33]. As a result, the final microstructure can preserve a stable stepped, semi-coherent interface that is both a product of growth selection and of *in situ* high-temperature annealing during the printing. HAADF intensity area profile across the interface (Figure 1h) indicate a 0.5 to 1 nm Ni-rich concentration gradient on the Cr-rich (BCC) side of the FCC/BCC boundary, in line with interfacial segregation confined to a few atomic planes. To further examine the origin of the gradual HAADF intensity increase on the Cr-rich side of the interface, a simplified single-row atomic-column model with ~20 nm thickness was constructed across the Cr–Ni interface and simulated using the abTEM algorithm [34]. The model consisted of progressively increasing Ni occupancy from pure Cr to pure Ni, represented by pure Cr, Cr–0.25Ni, Cr–0.50Ni, Cr–0.75Ni, and pure Ni atomic columns. The simulated HAADF image (Figure 1i) shows a monotonic increase in column intensity with increasing Ni content, consistent with the experimentally observed linear intensity gradient near the interface. This interpretation is further supported by STEM-EDX mapping of the same gradient region, where the Ni signal increases across the Cr-rich side approaching the FCC lamella (Figure 1j). Together, the simulated HAADF contrast and local EDX analysis indicate that the interfacial intensity gradient reflects near-interface Ni enrichment on the Cr-rich BCC side. In a semi-coherent K–S type interface, the boundary contains steps and misfit dislocations whose strain fields act as strong solute traps, and Ni enrichment can further reduce interfacial energy and locally relax misfit strain [35]. This near-interface chemistry can be explained by the processing path: rapid solidification promotes solute trapping, and subsequent cooling provides enough short-range diffusion for Ni to redistribute to these interfacial defect sites while remaining confined because Ni has limited equilibrium solubility in Cr-rich BCC away from the interface [36, 37].

### ***3.2. Nanomechanical testing***

Direct-pull *in situ* SEM microtensile testing was first performed to evaluate the strength–plasticity response of the ultrafine Cr–Ni eutectic. Arc-melted pure Cr and pure Ni specimens with comparable dog-bone geometry were also tested as reference materials to isolate the role of the ultrafine eutectic heterostructure. Figure 2a1 presents representative true stress–true strain profiles of the three systems, color-coded as Cr–Ni ultrafine eutectic (red), pure Cr (blue), and pure Ni (green). For the Cr–Ni eutectic, three microtensile tests were performed with maximum nominal strain targets of 15%, 25%, and 35%; the true stress–true strain curve from the specimen deformed to the largest nominal strain is shown in Figure 2a1, and the corresponding engineering stress–strain curves from all three tests are provided in the Supplementary Material. The engineering stress– strain curves show reproducible peak stresses, giving an ultimate tensile strength of $1110 \pm 30$ MPa. The representative true stress–true strain curve of the Cr–Ni eutectic shows a 0.2% offset yield stress of ~900 MPa, a maximum tensile stress of ~1.25 GPa, and sustained true plastic strain of at least ~24% before the test was interrupted prior to complete fracture. In comparison, pure Cr yields at ~400 MPa and reaches a maximum tensile stress and plastic strain at the onset of failure of ~700 MPa and 17%, respectively, whereas pure Ni yields at ~250 MPa and reaches a maximum tensile stress and plastic strain at the onset of failure of ~400 MPa and 19%, respectively. The initial linear slope of the engineering curves should not be interpreted as the elastic modulus, because strain was inferred from gripper displacement rather than measured directly from the gauge section and therefore includes contributions from system compliance and local alignment effects. The serrations observed in the Cr–Ni eutectic and pure Ni stress–strain curves are attributed to intermittent localized plastic flow, including discrete slip bursts and shear-band propagation, during displacement-controlled microscale tensile deformation.

It is important to note that the pure Cr and pure Ni reference specimens were prepared from arc-melted coarse-grained materials, with grain sizes typically exceeding 200 μm; therefore, the microscale tensile responses represent local small-volume behavior rather than bulk tensile properties [38]. The higher apparent strengths in pure metal samples arise from well-known small-volume effects: the gauge section samples a single crystallographic region, contains fewer pre-existing critical flaws, and restricts the operation of long-range dislocation sources, thereby increasing the stress required for plastic flow [39]. Similarly, the enhanced strain to failure, particularly in pure Cr, can be attributed to the reduced probability of sampling a fatal cleavage flaw and the limited volume available for crack nucleation and propagation [40, 41, 42].

Thus, the pure Ni and pure Cr data should be interpreted as local reference responses measured under the same microscale testing geometry, rather than as direct substitutes for conventional bulk tensile properties. For comparison, bulk polycrystalline Ni typically exhibits room temperature yield strength of ~60–150 MPa, ultimate tensile strength of ~300–450 MPa, and tensile elongation of ~30–50%, whereas bulk polycrystalline Cr has reported yield strength of ~350 MPa and tensile/fracture strength of ~400 MPa but very limited room-temperature ductility, often <2–3% elongation [43, 44]. In contrast, the EPBF processed Cr–Ni eutectic contains an ultrafine interlamellar spacing of ~450 nm and a colony size of only a few micrometers, such that the same gauge volume samples multiple lamellae, interfaces, and colony-scale features. The measured tensile response of the eutectic is, therefore, more representative of the composite microstructure within the microscale specimen, rather than being dominated by the response of a single coarse grain.

The work-hardening response further supports the superior tensile stability of the ultrafine eutectic. Figure 2a2 shows the work-hardening rate (WHR, $\theta = \frac{d\sigma}{d\varepsilon}$) as a function of true strain, where the Cr–Ni eutectic maintains a positive WHR over a wider strain range than both pure Cr and pure Ni. The representative postmortem morphologies (refer to the inset micrographs) reveal distinct failure modes among the three systems when WHR abruptly drops below zero due to strain localization and damage accumulation. Pure Cr fails by brittle crack propagation, with no clear evidence of stable shear-band formation prior to fracture. This behavior suggests that, once a critical crack nucleates in monolithic Cr, the available crack-tip plasticity is insufficient to blunt the crack or redistribute the local stress, leading to rapid failure. Pure Ni, in contrast, exhibits ductile localization characterized by shear offset and cup-and-cone-like fracture morphology, consistent with extensive plastic flow before final rupture. The Cr–Ni eutectic displays a fundamentally different response from pure Cr: instead of immediate catastrophic failure following crack initiation in the Cr-rich phase, deformation localizes through translamellar shear bands that cut across the lamellar heterostructure. This indicates that the Cr-rich BCC lamellae remain mechanically coupled to the surrounding Ni-rich FCC phase and participate in strain accommodation before final failure. The coupled stress–strain and WHR plots in Figure 2a3 identify the Considère crossover condition, where the WHR becomes equal to the true stress (i.e., $\frac{d\sigma}{d\varepsilon} = \sigma$) and consistently decreases beyond that point, corresponding to the onset of plastic instability [45]. This crossover is delayed to ~13.3% true strain and ~1.25 GPa in the Cr–Ni eutectic, compared with ~8.2% and ~0.70 GPa for pure Cr and ~6.3% and ~0.40 GPa for pure Ni. The delayed crossover indicates that the eutectic retains a higher work-hardening capacity before the onset of necking/localization, consistent with progressive strain accommodation within the ultrafine lamellar microstructure. Representative *in situ* SEM snapshots of the Cr–Ni dog-bone specimen (Figure 2b) show stable deformation up to large strain, followed by localized shear-band formation near final failure. Supplementary Video 1 presents a representative *in situ* tensile test of the Cr–Ni eutectic (deformed up to ~35% nominal strain) in an SEM. Supplementary videos 2 and 3 correspond to the tensile tests of reference pure Cr and pure Ni samples, respectively. *Postmortem* SEM imaging using both topographic and compositional contrast (Figure 2c1) reveals that these shear bands traverse the Ni-rich FCC and Cr-rich BCC lamellae rather than being confined to the softer phase or along the interface. Furthermore, comparison of the Cr-rich lamellae before and after deformation (Figure 2c2) shows measurable lamellar shape change, indicating that the Cr-rich BCC phase does not behave as a brittle reinforcement during tensile loading. Collectively, these observations indicate that the high tensile ductility of the Cr–Ni eutectic arises from plastic co-deformation of both constituent phases, rather than from isolated plastic flow of the Ni-rich phase alone.

While the tensile experiments establish an exceptional strength–plasticity response of the ultrafine Cr–Ni eutectic under a failure-sensitive loading condition, the highly localized deformation and final fracture of microtensile specimens make postmortem phase-resolved TEM analysis challenging. In addition, compression minimizes crack-opening-driven load drops and intermittent tensile shear-band bursts, producing a smoother work-hardening response that is more suitable for identifying deformation-stage transitions. Therefore, *in situ* micropillar compression was performed as a complementary nanomechanical test to impose larger plastic strains in a stable manner and to generate sufficient deformed volume for site-specific TEM lift-out. Unlike tension, compression suppresses catastrophic crack opening and allows the lamellar microstructure to be driven to higher strains, thereby providing a more suitable configuration for identifying the deformation mechanisms responsible for plastic co-deformation. The compression experiments were therefore used primarily to develop the *postmortem* microstructural basis for the tensile strength–plasticity response described above.

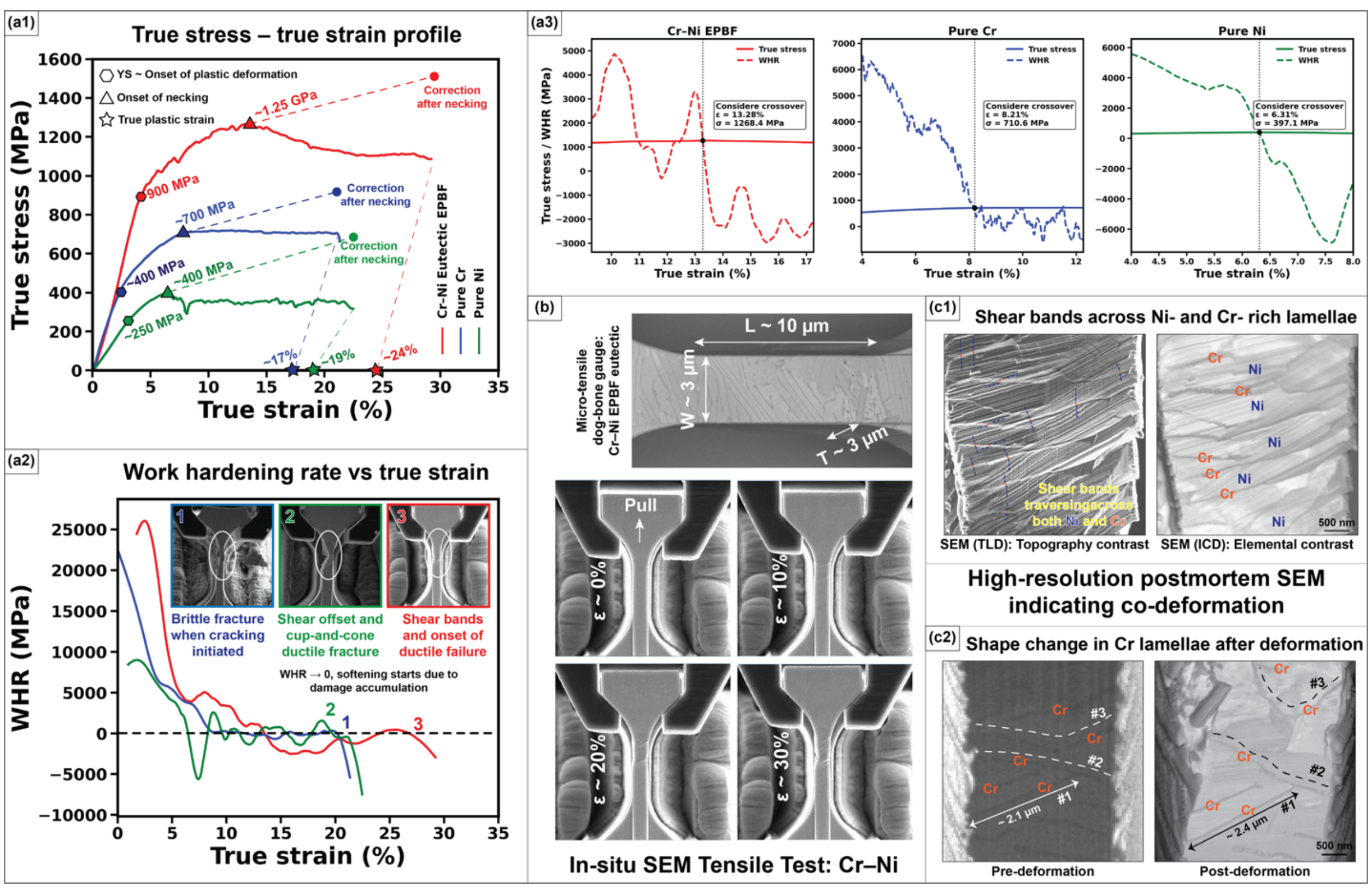

Figure 2. *In situ* SEM microtensile response of the EPBF Cr–Ni ultrafine eutectic compared with pure Cr and pure Ni reference specimens. (a1) True stress–true strain profiles of the Cr–Ni eutectic, pure Cr, and pure Ni, showing the superior strength–plasticity combination of the eutectic. The onset of plastic deformation, maximum true tensile stress/onset of necking, and final true plastic strain are marked for each specimen. Dashed extensions indicate post-necking corrected true-stress estimates. (a2) Work-hardening rate (WHR) as a function of true strain for the three systems, showing delayed softening and tensile instability in the Cr–Ni eutectic. Representative *postmortem* morphologies identify brittle cracking in pure Cr, shear offset/cup-and-cone-type ductile fracture in pure Ni, and shear-band-mediated failure in the Cr–Ni eutectic. (a3) True stress and WHR plotted against true strain for each material, identifying the Considère crossover condition where WHR equals the true stress. (b) Representative *in situ* SEM snapshots of the Cr–Ni microtensile dog-bone specimen during direct-pull testing, showing stable deformation up to large strain before localization. (c1) High-resolution postmortem SEM images acquired using topographic contrast and compositional contrast, revealing shear bands that traverse both Ni-rich and Cr-rich lamellae. (c2) Comparison of Cr-rich lamellae before and after deformation, showing measurable lamellar shape change and indicating plastic participation of the Cr-rich phase during tensile deformation.

*In situ* SEM compression tests exhibit a notable combination of high strength ($\sigma_c$ ~1.55 GPa) and plastic strain ($\varepsilon_p$ ~32%) (Figure 3a). The 0.2% offset yield strength ($\sigma_y$) is ~950 MPa, similar to that measured under tension, followed by pronounced early work hardening and sustained hardening to large strain. It should be noted that the yield strength in our study is significantly higher than the study by Kossowsky *et al*., that reported $\sigma_y$ ~ 530 MPa for a lamellar eutectic Cr–Ni microstructure [27]. However, the lower $\sigma_y$ in their study could be attributed to coarser microstructure with an average interlamellar spacing of 10 μm. The early work hardening is typical in heterostructures due to back-stress strengthening as a result of GNDs when there is plastic deformation incompatibility between disparate phases [46]. The corresponding Kocks–Mecking plot (inset; WHR, $\theta = \frac{d\sigma}{d\varepsilon}$ vs stress in plastic deformation regime) shows an initial rapid reduction in $\theta$ with increasing stress in stage II and a distinct inflection consistent with a transition into the stage III regime, where hardening typically continues to decrease toward failure as $\theta \rightarrow 0$ [47]. Notably, a second change in slope is observed at higher stress, indicating an additional transition that delays the approach to $\theta \approx 0$ and suggests activation of a different deformation process.

Sequential SEM snapshots acquired during loading (Figure 3b) confirm stable, uniform deformation without observable cracking, or brittle failure; consistent with this, the specimen geometry before and after deformation shows no obvious shear offset. The specimens maintained their cylindrical geometry throughout as confirmed by comparing specimen dimensions before and after compression and considering zero volume change in plastic deformation. Supplementary Video 4 displays a representative *in situ* compression test in an SEM. *Postmortem* TEM of foils lifted from the compressed specimen (Figure 3c1–c2) reveals continuous Cr-rich BCC lamellae without cracking even at high strain. Finally, correlative high-resolution SEM imaging using the through lens detector (topographic contrast; Figure 3d1) and in-column detector (compositional contrast; Figure 3d2) shows shear bands traversing both Ni-rich and Cr-rich lamellae, similar to the observation after tensile test, providing direct indication of plastic co-deformation of the two phases. The participation of the Cr-rich lamellae in shear-band-mediated deformation is unusual given the commonly reported brittle response of monolithic Cr at room temperature and its elevated ductile-to-brittle transition temperature [48]. Detailed microstructure characterization of each phase after deformation is therefore necessary to reveal the underlying deformation mechanisms.

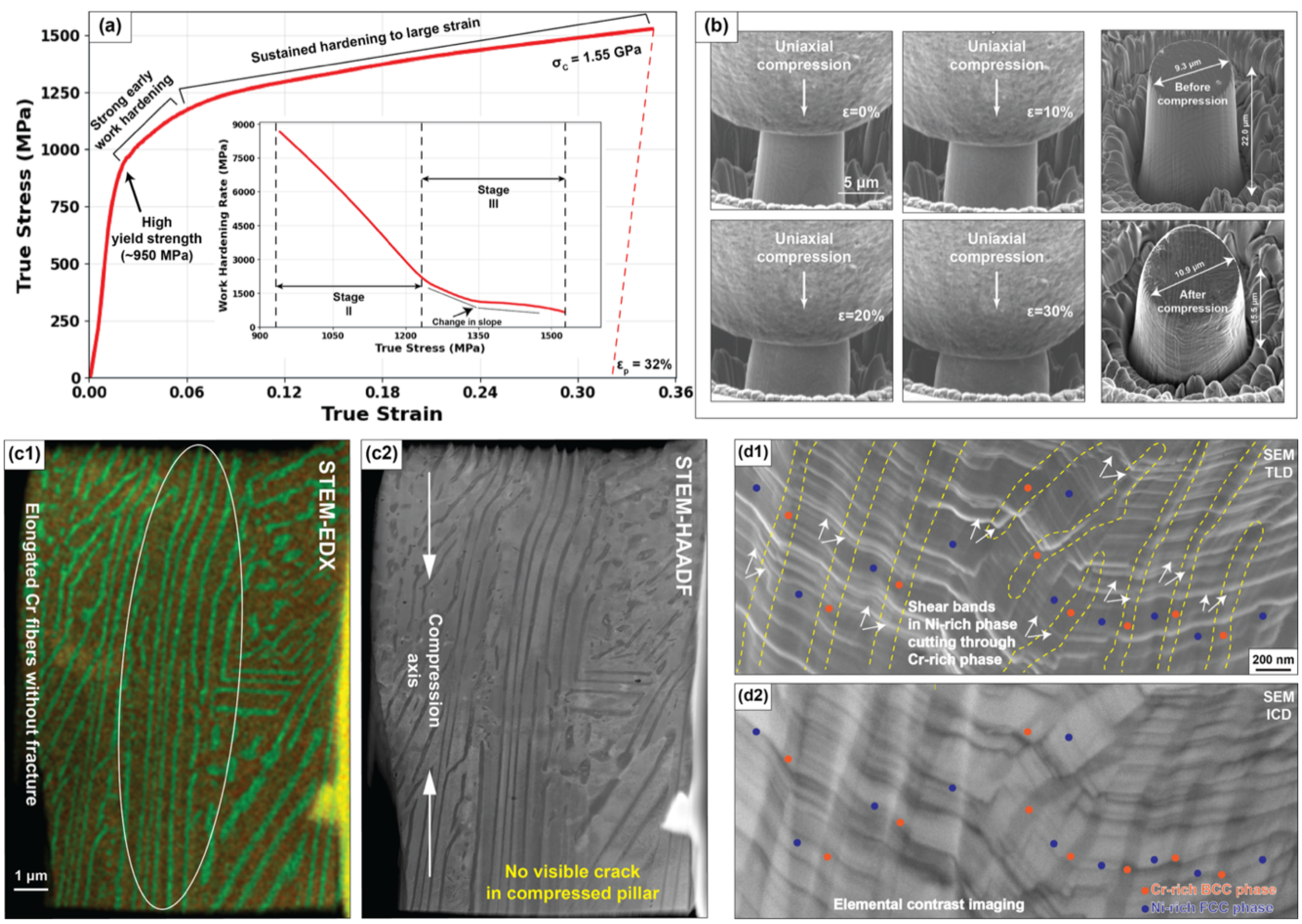

Figure 3: (a) True stress–true strain curve from *in situ* compression test showing $\sigma_y \sim 0.95$ GPa, $\sigma_c \sim 1.55$ GPa, and $\varepsilon_p \sim 32\%$; inset: Kocks–Mecking plot indicating distinct hardening regime transitions. (b) Time-resolved SEM snapshots at selected strains and specimen geometry before/after compression, demonstrating stable, uniform deformation without catastrophic failure. (c1–c2) *Postmortem* STEM-EDX and HAADF from the compressed specimen showing intact, elongated Cr-rich BCC lamellae without visible cracking. (d1-d2) Correlative SEM TLD/ICD imaging highlighting shear bands that traverse both Ni-rich FCC and Cr-rich BCC lamellae, suggesting plastic deformation of both phases.

### *3.3. Deformed microstructure*

To resolve phase-level strain accommodation after large plastic deformation, *postmortem* TEM was performed on each constituent phase. We first describe deformed microstructures in the Ni-rich FCC lamellae, followed by those in the Cr-rich BCC lamellae. Two-beam bright-field/dark-field imaging under $\vec{g}\,[1\,1\,\bar{1}]$ (Figure 4a1–a2) reveals a high density of lattice dislocations in the Ni-rich phase after compression, together with well-defined slip traces and deformation twins confined to a specific $\{1\,1\,1\}$ plane. Complementary two-beam imaging under $\vec{g}\,[0\,\bar{2}\,2]$ (Figure 4b1–b2) provides enhanced visibility of the dislocation network and indicates the operation of multiple slip systems; this reflection satisfies the invisibility criterion for a larger subset of FCC $^{a}/_{2}\langle 0\,1\,1\rangle$ Burgers vectors (capturing 5 of the 6 $\langle 0\,1\,1\rangle$ variants), whereas $\vec{g}\,[1\,1\,\bar{1}]$ is sensitive to only 3 out of 6. The inset HAADF image in Figure 4b2 resolves an individual dislocation core at atomic resolution, and Burgers circuit analysis identifies a perfect dislocation with $b = {}^{a}/_{2}\ [0\ \bar{1}\ 1]$, consistent with dislocation glide on $\{1\ 1\ 1\}\ \langle 0\ 1\ 1\rangle$ systems in the Ni-rich FCC lamellae.

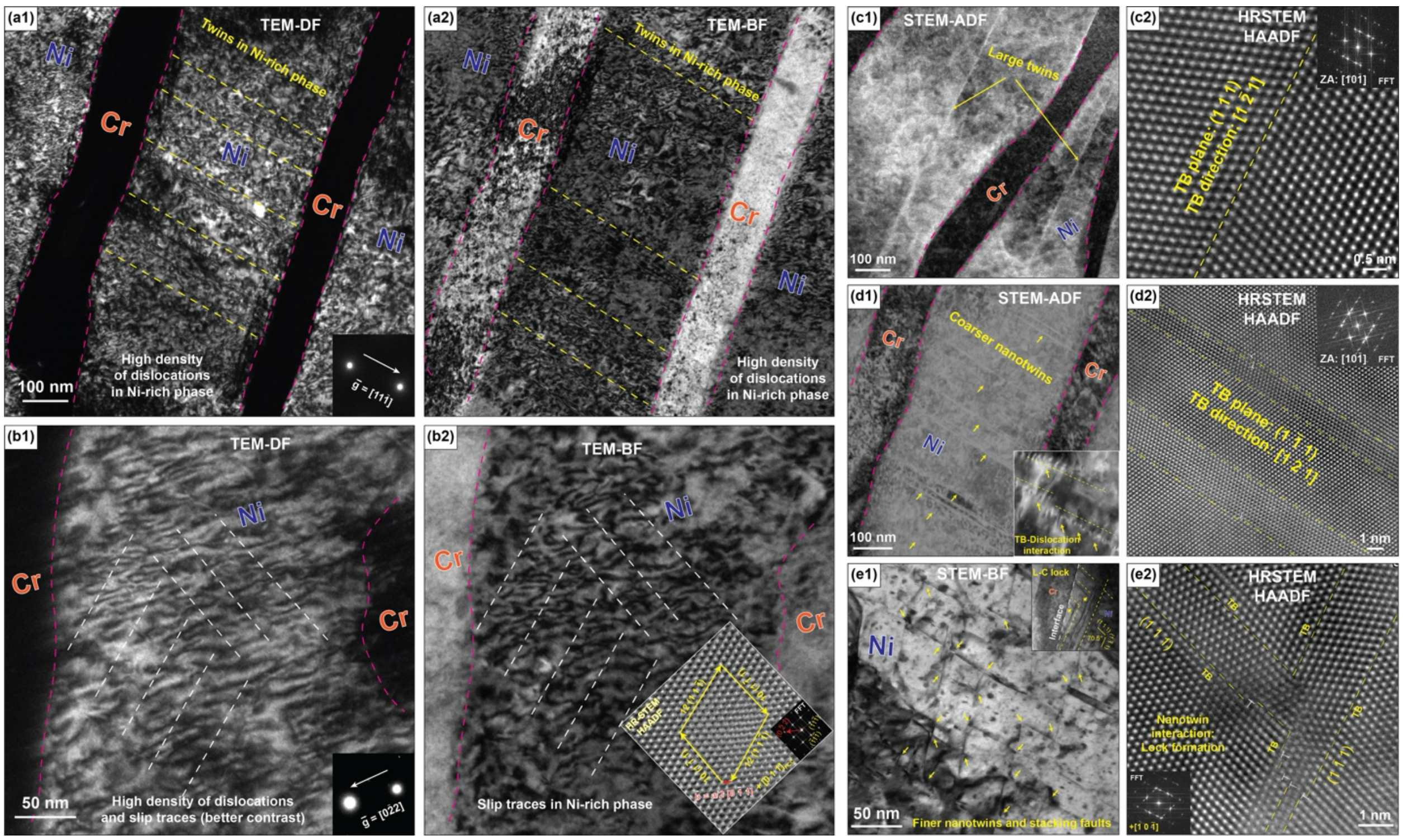

Figure 4: (a1–a2) Two-beam TEM dark-field/bright-field images ($\vec{g}\ [1\ 1\ \bar{1}]$) showing a high dislocation density in the Ni-rich phase, together with slip traces and deformation twins within a single lamella bounded by Cr-rich lamellae. (b1–b2) Two-beam TEM imaging ($\vec{g}\ [0\ \bar{2}\ 2]$) reveals the dislocation network and multiple slip traces with improved contrast; inset in (b2) shows a HAADF image of an individual dislocation core and Burgers-circuit analysis identifying a perfect FCC dislocation $b = {}^{a}/_{2}\ [0\ \bar{1}\ 1]$. (c1–c2) ADF and HAADF images showing coarse deformation twins with atomically resolved twin boundaries in the Ni-rich lamella. (d1–d2) Additional ADF and HAADF images highlighting coarser nanotwins and their boundary structure. (e1–e2) BF STEM and HAADF images showing finer nanotwins, twin–twin interactions, and lock formation at twin junctions, indicating strong twin-mediated hardening in the Ni-rich FCC phase.

Post-deformation on-zone STEM imaging of the Ni-rich FCC lamellae reveals a hierarchical twin-mediated deformation substructure superimposed on the high dislocation density described above. In addition to dense $\{1\ 1\ 1\}\ \langle 0\ 1\ 1\rangle$slip activity, STEM and HR-STEM images (Figure 4c–d) show both coarse [~1–5 µm long with spacing of ~100–500 nm] deformation twins and finer nanotwins [~100–500 nm long with spacing of ~1–5 nm] within individual Ni-rich lamellae, indicating progressive activation of twinning as strain increases. The atomically resolved twin boundaries are planar and faceted, and their crystallography is consistent with FCC deformation twins on $\{1\ 1\ 1\}$planes along $\langle 1\ 2\ 1\rangle$ directions (i.e., coherent Σ3-type twin interfaces) [49]. Figure 4e1-e2 further shows twin–twin intersections and junctions, where interacting twinning partials leave sessile residual defects (Lomer-Cottrell lock formation), creating additional obstacles to both dislocation motion and twin boundary migration [50]. The coexistence of dense dislocations, multiple slip traces, and interacting nanotwins therefore indicates that the Ni-rich phase accommodates strain by a coupled slip–twinning mechanism rather than by dislocation glide alone. These observations provide a microstructural basis for the pronounced early work hardening and continued hardening to large strain observed in the micropillar compression response.

Post-deformation TEM of the Cr-rich BCC lamellae was performed to determine the plastic flow activity in the hard phase. Low-magnification STEM and TEM imaging show that the Cr-rich lamellae remain continuous and crack-free after micropillar compression, with no evidence of cleavage or lamellar fragmentation despite the large applied strain. An on-zone STEM-BF image (Figure 5a1) reveals a high density of deformation-induced dislocations in the Cr-rich phase, with an estimated (calculated from STEM image: total length of dislocation lines per unit volume) dislocation density of $\rho \approx 5 \times 10^{14}\ m^{-2}$. Under the [1 1 1] BCC zone-axis condition (Figure 5a2), the dislocation lines preferentially align with the projected $\langle 1\ 1\ 1 \rangle$ directions lying on the $\{0\ 1\ 1\}$ planes along real $\langle 1\ 2\ 1 \rangle$ directions, consistent with screw-dominated segments in BCC where line direction is parallel to Burgers vector.

To further assess the Burgers-vector character, two-beam BF-STEM imaging was performed under different diffraction conditions (Figure 5b1–b3). The representative dislocation lines remain visible for $\vec{g}$ [2 0 0] and $\vec{g}$ $[\bar{1}\ 1\ 0]$, where $\vec{g}.\vec{b} \neq 0$, but become invisible for $\vec{g}$ $[\bar{1}\ \bar{1}\ 0]$, where $\vec{g}.\vec{b} = 0$. This contrast behavior is consistent with $^{a}/_{2}\,\langle 1\ \bar{1}\ 1 \rangle$-type dislocations in the Cr-rich BCC phase. In addition, an on-zone BF-STEM image acquired from a highly strained region of the compressed TEM foil shows a nearly micrometer-long Cr-rich lamella segment containing a dense dislocation network (Figure 5d), confirming that dislocation activity in Cr is not limited to isolated local regions. A comparison of strain evolution in the pre- (Figure 5e1–e5) and post-deformed (Figure 5f1–f5) Cr lamellae has been presented using geometric phase analysis (GPA). The strain level is significantly higher after deformation. Plastic deformation of the Cr-rich lamellae is even clearer from the GPA as the vertical and horizontal strain components exhibit compressive and tensile strains, respectively. Under the absence of plastic deformation in Cr, the Cr-rich phase would exhibit elastic strain under compression and fracture once the acting tensile stress on the Cr lamellae exceeds its fracture strength.

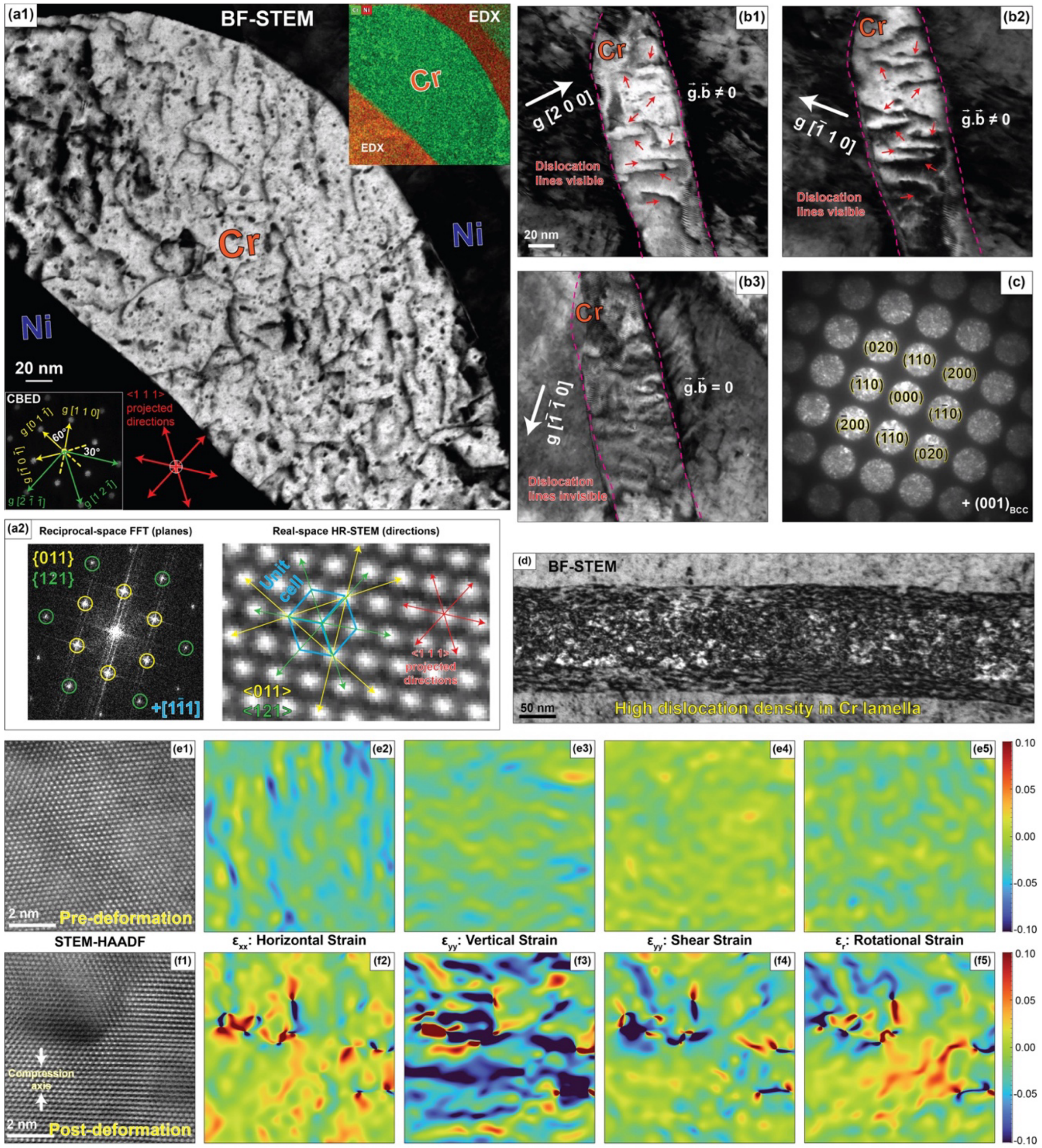

Figure 5: (a1) On-zone BF-STEM image showing a high density of deformation-induced dislocations in the Cr-rich phase; dislocation lines aligned with projected ⟨1 1 1⟩ directions are marked (inset image on top right is the EDX map used to verify thr Cr phase and the bottom left CBED pattern is used to identify the crystallographic directions). (a2) Corresponding BCC zone-axis diffraction pattern and real-space image used to establish the projected directions and indexing. (b1–b3) Two-beam condition BF-STEM images along different diffraction conditions, $\vec{g}$ [2 0 0], $\vec{g}$ [$\bar{1}$ 1 0], and $\vec{g}$ [$\bar{1}$ $\bar{1}$ 0], to identify Burgers vector by $\vec{g}.\vec{b}$ calculations and (c) corresponding [0 0 1] zone axis CBED pattern used to tilt the specimen along different $\vec{g}$. (d) On-zone BF-STEM image acquired from the highly strained region of the compressed TEM foil, showing Cr-rich BCC lamella containing a high density of deformation-induced dislocations over a nearly micrometer-long segment. (e1–e5) GPA-derived strain/rotation maps of the as-built Cr lamellae. (f1–f5) Corresponding GPA maps after deformation showing significantly increased strain levels and heterogeneous strain partitioning within the Cr-rich phase.

### *3.4. Evolution of deformed microstructure*

*Postmortem* TEM of the compressed specimen establishes the key deformation modes – dislocation slip and twinning in the Ni-rich FCC lamellae and pronounced dislocation activity in the Cr-rich BCC lamellae – yet it does not directly capture the sequence by which these microstructures develop. To resolve the evolution with increasing strain, nanoindentation was performed using a low peak load (~50 mN) such that the plastic zone remained comparable to the lifted cross-section lamella height (~3 μm). The corresponding indentation depth is ~500 nm; using a first-order estimate that the plastically affected depth extends to ~5x the indentation depth [51], the expected plastic zone size is ~2.5 μm. Figure 6a shows the indentation site and the associated load–displacement curve (inset), while the yellow rectangle denotes the cross-section lift-out region and Figure 6b shows the schematic of the extracted TEM lamella. The lamella exhibits a clear strain gradient from bottom to top, enabling division into three zones: Zone C (bottom) resembles the undeformed microstructure, Zone B (intermediate) shows plasticity primarily in the Ni-rich phase, and Zone A (near the indenter) exhibits co-deformation of both phases. In Zone B (Figure 6d1–d2), the Ni-rich lamellae contain a high density of forest dislocations (average dislocation density of $\rho \approx 8 \times 10^{14}\ m^{-2}$) [as measured from STEM image in Figure 6d2: total length of dislocation lines per unit volume], with dislocation accumulation concentrated near the Ni/Cr interfaces consistent with plastic incompatibility, whereas the Cr-rich lamellae show no visible evidence of slip. In Zone A (Figure 6c1–c2), the Ni-rich phase deforms predominantly by twinning in addition to dislocation glide, while the Cr-rich phase displays a high density of dislocations, indicating activation of plasticity at higher local strain. Collectively, these observations support a sequential deformation pathway in which Ni-rich slip governs initial yielding, followed by twin-assisted FCC plasticity and mobile dislocation activity in the harder Cr phase that together sustain work hardening and enable large compressive deformability.

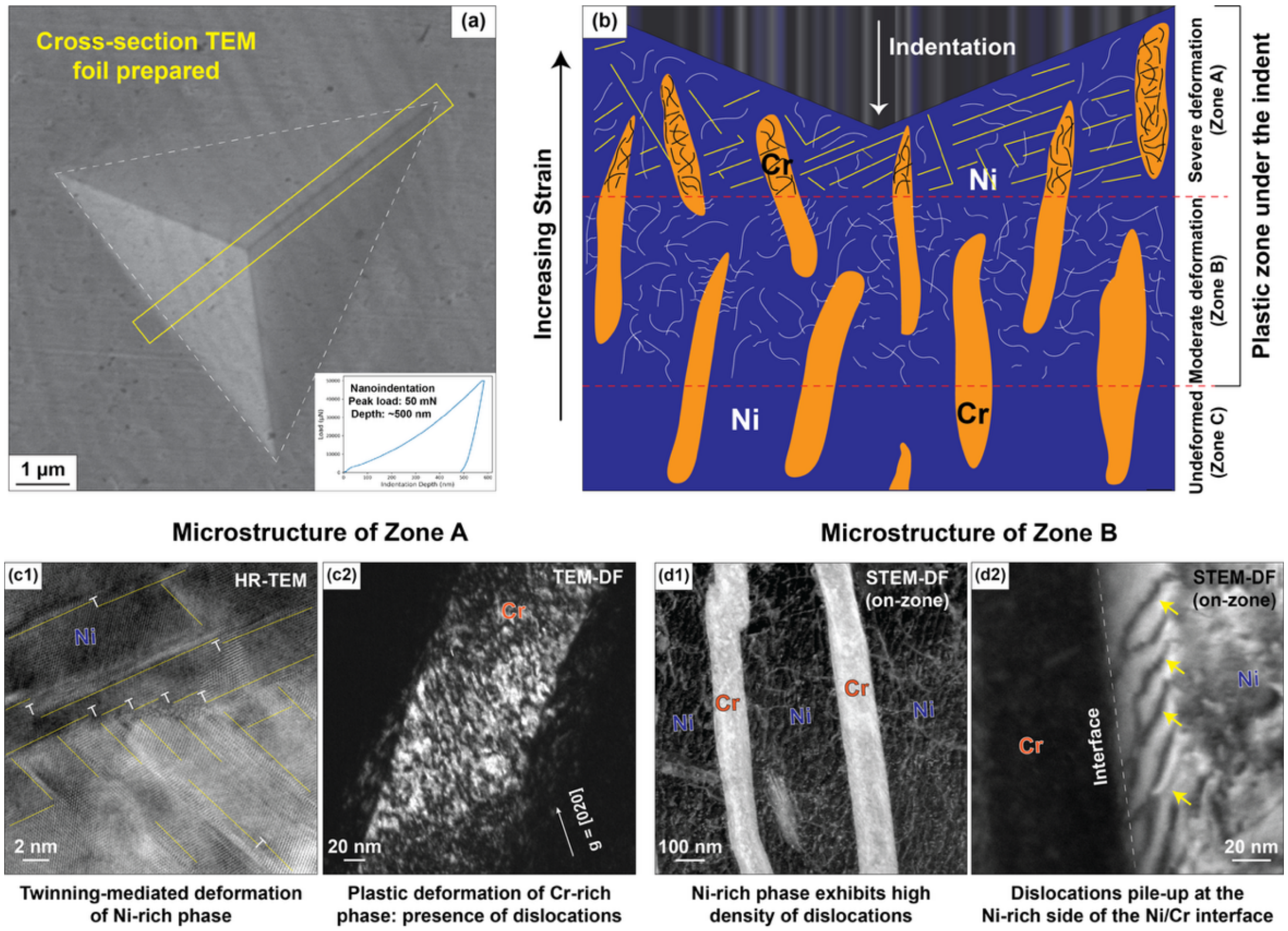

Figure 6: (a) SEM image of the indentation site with load–displacement curve (inset); yellow box indicates the cross-section TEM lift-out region. (b) Schematic of the extracted cross-section TEM lamella showing a through-thickness strain gradient from bottom (low strain) to top (high strain), enabling division into Zones C to A. (c1–c2) Zone A (highest strain): twinning-dominated deformation in Ni-rich FCC with high dislocation density in Cr-rich BCC, indicating phase co-deformation. (d1–d2) Zone B (intermediate strain): dense perfect dislocations in Ni-rich FCC concentrated near interfaces, with no visible evidence of plasticity in Cr-rich BCC.

## *4. Discussion*

Building on the as-printed and post-deformation microstructural evidence presented above, we now discuss the mechanistic origins of the high strength and sustained hardening – emphasizing phase-specific deformation in Ni and Cr, the enabling role of the stepped semi-coherent interface, and how the resulting plastic co-deformation governs tensile ductility. Therefore, the mechanistic discussion is organized into two parts: (i) the origin of the high yield strength, governed primarily by the onset of plasticity in the Ni-rich FCC lamellae within an ultrafine heterogeneous microstructure, and (ii) the origin of the sustained work hardening and large compressive deformability enabled by plastic co-deformation of both phases – manifested as deformation twinning in the Ni-rich phase and pronounced dislocation activity in the Cr-rich BCC phase. Accordingly, we first analyze the strengthening contributions that set the yield strength and then develop a phase-resolved picture of the subsequent deformation sequence and interface-mediated mechanisms that delay damage accumulation and suppress microstructural failure to large plastic strain.

### *4.1. High yield strength*

The high yield strength of the ultrafine Cr–Ni eutectic cannot be interpreted as a simple average of monolithic pure Ni and pure Cr responses. Although the pure Ni and pure Cr microtensile tests provide useful baseline references, the actual deforming FCC phase in the eutectic is a Cr-containing Ni-rich solid solution, and its strength is further amplified by ultrafine lamellar confinement, elastic–plastic incompatibility, and strain-gradient-driven dislocation storage at the FCC/BCC interfaces. The yield strength ($\sigma_y$), at small plastic strain offset of ≈0.2%, with the assumption of plastic deformation confined in Ni phase and elastic loading of Cr, can be estimated as the weighted average of stresses in the respective phases:

$$\sigma_y = \sigma_{Ni}V_{Ni} + \sigma_{Cr}V_{Cr} \quad (1)$$

where $\sigma_{Ni}$ and $\sigma_{Cr}$ present the strengthening contribution from Ni and Cr, and $V_{Ni}$ and $V_{Cr}$ are their respective volume fractions, ~0.7 and 0.3, respectively (refer to Table 1). The Ni-rich FCC phase, being softer than the Cr-rich BCC phase at room temperature, is expected to yield first while the Cr-rich phase still loads predominantly elastically. This elastic–plastic mismatch imposes a strain gradient across the Ni/Cr interfaces, such that a high density of interface-associated dislocations (dominated by GNDs) is required to maintain compatibility. These dislocations can glide and multiply within the confined Ni-rich lamellae and contribute to the early strengthening response through dislocation storage and pile-up at the interface (Figure 6d1–d2) [52, 53]. As a first-order estimate of the dislocation contribution, the flow stress increment associated with dislocation interactions can be expressed using Taylor hardening:

$$\sigma_f = M\alpha\mu b\sqrt{\rho_{GND+SSD}} \quad (2)$$

Where $M$ is Taylor factor ($\approx 3$ for FCC), $\alpha$ is a geometric factor ($\approx 0.3$ typical for forest interaction in FCC) [54], $\mu$ is the shear modulus, $b$ is the Burgers vector of a perfect $a/2\ \langle 0\ 1\ 1\rangle$ dislocation (refer to Figure 4b2) in Ni ($\approx 0.249\ nm$), $\rho$ is the dislocation density (considering both geometrically necessary and statistically stored dislocations) measured in the lowest-strain regime of the Ni-rich phase within plastic zone ($\approx 8 \times 10^{14}\ m^{-2}$ from Figure 6d1). The dislocation density was estimated using:

$$\rho_{GND+SSD} = \frac{L}{l \times w \times t} \quad (3)$$

where, $L$ is the total visible dislocation-line length within the field of view, $l$ and $w$ are the length and width of the field of view, and $t$ is the TEM foil thickness. The foil thickness was estimated to be $\approx 50\ nm$ from FIB measurements, with an uncertainty of approximately $\pm 10\ nm$ corresponding to an uncertainty of ($\approx 20\%$) in the dislocation-density estimate. This value is used as a representative order-of-magnitude estimate of early strain-gradient-mediated dislocation storage in the Ni-rich lamellae. Directly measuring the dislocation density associated with the onset of plastic deformation in tensile or compression specimens is experimentally challenging because the relevant deformation volume is difficult to isolate after global loading. In contrast, the shallow-indentation experiment produces a localized and well-defined strain gradient beneath the indent, allowing the low-to-intermediate strain region to be identified and examined by site-specific TEM. Therefore, the dislocation density measured beneath the indent is used as a mechanistic estimate of early interface-associated dislocation storage, rather than as an exact measurement from the tensile or compression specimens. In addition, the local GND density near FCC/BCC interfaces may exceed the field-of-view-averaged value because dislocation storage is spatially concentrated at interfaces where plastic incompatibility is accommodated.

Because the Ni-rich phase contains substantial Cr ($\approx 36\ wt.\%$, i.e., $\approx 39\ at.\%$) dissolved, an effective shear modulus was estimated using the Voigt rule of mixtures [55]:

$$\mu_{eff} = C_{Ni}\mu_{Ni} + C_{Cr}\mu_{Cr} \tag{4}$$

As quantified from STEM-EDX measurements, values of atomic fractions, $C_{Ni}$ and $C_{Cr}$, are 0.61 and 0.39, respectively. Respective values of $\mu_{Ni}$ and $\mu_{Cr}$ have been reported as 76 GPa and 115 GPa [56]. Therefore, a concentration-weighted average of the elemental shear moduli gives $\mu_{eff} \approx 91\ GPa$ for the Ni-rich solid solution, which is consistent with the reported $G$ values for concentrated Ni(Cr) solid solutions derived from first principles calculations [57]. Substituting these parameters into *Equation (2)* yields $\sigma_f \approx 575\ MPa$.

Another significant contribution comes from solid solution hardening owing to substantial Cr ($C_{Cr} \approx 0.4$) dissolved in Ni. Here we invoke the Varvenne-Curtin model [57] of solid solution strengthening (SSS) for FCC alloys to quantify the strength coming from SSS. Although originally developed for FCC multicomponent alloys, the Varvenne–Curtin framework is also applicable to binary concentrated FCC solid solutions [58]. Table 2 summarizes the parameters used for the model.

Table 2: Parameter details used in the Varvenne-Curtin model

| **Quantity** | **Symbol** | **Value used** | **Comment** |
|---|---|---|---|
| Ni atomic fraction | $C_{Ni}$ | $\approx 0.60$ | From Ni–40 at.% Cr |
| Cr atomic fraction | $C_{Cr}$ | $\approx 0.40$ | From Ni–40 at.% Cr |
| Apparent atomic volume of Ni | $V_{Ni}$ | $\approx 10.94$ Å$^3$ | FCC apparent atomic volume |
| Apparent atomic volume of Cr | $V_{Cr}$ | $\approx 12.27$ Å$^3$ | FCC apparent atomic volume |
| Shear modulus | $\mu$ | $\approx 90\ GPa$ | Assumed average shear modulus |
| Poisson's ratio | $\upsilon$ | $\approx 0.30$ | Assumed average alloy value |
| FCC line-tension parameter | $\alpha$ | $\approx 0.125$ | Standard FCC VC-model choice |
| Taylor factor, polycrystal | $M$ | $\approx 3$ | FCC random polycrystal |
| Room temperature | $T$ | 300 K | Room-temperature estimate |
| Reference strain rate | $\dot{\varepsilon}_0$ | $10^4/s$ | VC thermal activation term |
| Applied strain rate order | $\dot{\varepsilon}$ | $5 \times 10^{-3}/s$ | Representative test rate |
| Boltzmann constant | $k_B$ | $8.6 \times 10^{-5} eV/K$ | Used in thermal correction |

The average atomic volume is calculated using a rule of mixtures:

$$\bar{V} = C_{\mathrm{Ni}}V_{\mathrm{Ni}} + C_{\mathrm{Cr}}V_{\mathrm{Cr}} \tag{5}$$

The misfit volume of each element is defined as:

$$\Delta V_i = V_i - \bar{V} \tag{6}$$

Concentration-weighted mean-square misfit volume, i.e., the collective misfit term is given as:

$$\sum_i C_i\, \Delta V_i^2 = C_{\mathrm{Ni}}\Delta V_{\mathrm{Ni}}^2 + C_{\mathrm{Cr}}\Delta V_{\mathrm{Cr}}^2 \tag{7}$$

Substituting the values from Table 2 results in:

$$\sum_i C_i\, \Delta V_i^2 = 0.425\ \text{Å}^6 \tag{8}$$

For an FCC lattice, the Burgers vector is:

$$b = \frac{a}{\sqrt{2}} \tag{9}$$

The lattice parameter can be estimated from the average atomic volume using:

$$a = (4\bar{V})^{1/3} \tag{10}$$

Therefore:

$$b = \frac{(4\bar{V})^{1/3}}{\sqrt{2}} \tag{11}$$

Substituting the average atomic volume:

$$b = \frac{(4\times 11.472)^{1/3}}{\sqrt{2}} = 2.53\text{Å} \tag{12}$$

The dimensionless misfit-volume term in the Varvenne-Curtin model is:

$$\frac{\sum_i C_i \Delta V_i^2}{b^6} \tag{13}$$

Using the afore-obtained collective misfit term and Burgers vector leads to:

$$\frac{\sum_i C_i \Delta V_i^2}{b^6} = \frac{0.425}{263.2} \approx 1.6 \times 10^{-3} \tag{14}$$

Now, the model estimates the athermal critical resolved shear stress increment as:

$$\Delta\tau_{y0} = 0.01785\alpha^{-1/3}\bar{\mu}\left(\frac{1+\bar{\nu}}{1-\bar{\nu}}\right)^{4/3}\left[\frac{\sum_i C_i \Delta V_i^2}{b^6}\right]^{2/3} \tag{15}$$

Substituting the values gives $\Delta\tau_{y0} \approx 104$ MPa. This is the athermal, or 0 K, solid-solution strengthening contribution in resolved shear stress. To estimate the same at room temperature, we need to incorporate a thermal activation correction [58]. The corrected resolved shear stress is given as:

$$\Delta\tau_y(T) = \Delta\tau_{y0}\left[1 - \left(\frac{k_B T}{\Delta E_b}\ln\frac{\dot{\varepsilon}_0}{\dot{\varepsilon}}\right)^{2/3}\right] \quad (16)$$

Where $\Delta E_b$ is energy barrier defined as:

$$\Delta E_b = 1.5618\alpha^{1/3}\bar{\mu}b^3\left(\frac{1+\bar{\nu}}{1-\bar{\nu}}\right)^{2/3}\left[\frac{\sum_i c_i \Delta V_i^2}{b^6}\right]^{1/3} \quad (17)$$

Substituting all the parameters with their values from Table 2 yields: $\Delta\tau_y(300\text{ K}) \approx 60$ MPa. The Varvenne-Curtin model gives the strengthening increment in resolved shear stress. For comparison with a polycrystalline or constrained micropillar compression stress, the resolved shear stress increment can be converted to a macroscopic stress increment using the Taylor factor:

$$\Delta\sigma_{ss} = M\Delta\tau_y \approx 180\text{ MPa} \quad (18)$$

Eutectic colony boundaries, i.e., Ni–Ni grain boundaries, are likely secondary strengthening sources since the micropillars sample multiple such colonies, that could be directly quantified following the Hall-Petch equation for grain boundary strengthening:

$$\sigma_{H-P} = \sigma_0 + kd^{-1/2} \quad (19)$$

For simplicity, we can assume the Hall-Petch parameters to be close to that of pure Ni, where the friction stress ($\sigma_0$) and Hall-Petch coefficient ($k$) are often approximated as ~50 MPa and 0.25 MPa/m$^{1/2}$, respectively [59], which results in $\sigma_{H-P} \approx 160$ MPa for an average eutectic colony size of ~5 μm (Figure 1a). Note that the eutectic colony size is significantly larger than the eutectic interlamellar spacing, so the estimated strengthening contribution from Ni–Ni grain boundaries is relatively smaller.

The total stress in Ni is given by the sum of equations (2), (18) and (19):

$$\sigma_{Ni} = \sigma_f + \Delta\sigma_{SS} + \sigma_{H-P} \approx 915\text{ MPa} \quad (20)$$

Since the postmortem microscopy has shown Cr deforms predominantly elastically during initial deformation (very low strained regime after nanoindentation), $\sigma_{Cr}$ can be approximated as:

$$\sigma_{Cr} = E_{Cr} \times \varepsilon \quad (21)$$

At 0.2% offset yield, this becomes ~580 MPa considering the elastic modulus of Cr ($E_{Cr}$) to be 290 GPa [60]. Substituting the individual stress components along with respective volume fractions in equation (1) produces the modeled yield strength of ~815 MPa. It should also be noted that equation (1) assumes ideal iso-strain loading of continuous lamellae oriented parallel to the uniaxial loading axis. Because the lamellae in the actual EPBF microstructure are arranged within colonies with varying orientations, rather than being uniformly parallel to the tensile/compressive axis (Figure 1a), deviations from ideal iso-strain load sharing may also contribute to the discrepancy between the measured and calculated yield strengths.

The remaining difference between the modeled and measured (~900–950 MPa) yield strengths suggests an additional interface-mediated strengthening contribution that is not fully captured by solid-solution strengthening, Taylor dislocation strengthening, or elastic load sharing alone. The Taylor relation using the TEM-estimated total dislocation density captures the forest-strengthening contribution from both statistically stored and geometrically necessary dislocations. Therefore, a separate GND-based strengthening term was not added to avoid double counting. However, this scalar dislocation-density estimate does not fully account for long-range hetero-deformation-induced back stresses arising from elastic–plastic incompatibility between the Ni-rich FCC and Cr-rich BCC lamellae [52], nor does it capture source-limited plasticity imposed by the ultrafine lamellar geometry since the FCC/BCC interfaces restrict the available dislocation source length and slip distance, thereby increasing the stress required to initiate and sustain plastic flow [61]. These interface-mediated contributions are therefore expected to account for part of the remaining difference between the modeled and measured yield strengths. The possible contribution from source-limited plasticity can be understood from the scaling [62]:

$$\sigma_{source} = M\tau_{source} = \frac{MA\mu b}{L} \tag{22}$$

where L is the available dislocation source length and A~1 as an order of magnitude prefactor. Taking (L) to be comparable to the Ni-rich lamella thickness (~350 nm) gives a source-operation stress on the order of 100 MPa. Addition of equation (22) to (20) results in calculated yield strength that matched well with the experimental measurement, suggesting that confined-layer source limitation can plausibly contribute to the residual interface-mediated strengthening. However, the effective source length and interfacial boundary conditions are not directly measured, so this contribution is treated as a qualitative scaling argument rather than as a separate additive strengthening term in equation (20). The slightly higher yield strength measured in compression may also arise from the larger sampled volume of the compression specimen, which contains a greater number of eutectic colonies and Ni(Cr) FCC grain/colony boundaries. As a result, the Hall–Petch-type contribution from the Ni(Cr) FCC boundaries, estimated in Equation (19), may be more fully represented in compression than in the smaller microtensile gauge volume.

### 4.2. Sustained deformation up to large plastic strain

Beyond initial yielding, the persistence of work hardening to large plastic strain indicates a transition from soft-phase-dominated flow to a regime where additional deformation mechanisms are gradually activated. Our *postmortem* TEM and strain-gradient nanoindentation results suggest a clear sequence: at lower strains, plasticity is largely confined to the Ni-rich FCC lamellae via dislocation glide, whereas at higher strains the Ni-rich phase increasingly accommodates deformation by twinning while the Cr-rich BCC lamellae develop substantial dislocation activity. This phase-by-phase evolution is expected to fundamentally change the hardening capacity, because twins subdivide the FCC lamellae and introduce strong internal barriers, while activation of BCC slip in Cr redistributes load and suppresses premature fracture of the hard phase. In the following, we first discuss (i) the origin and hardening contribution of deformation twinning in the Ni-rich phase, and then (ii) analyze the origin of plasticity in the Cr-rich phase – highlighting the roles of length scale, stepped semi-coherent interface structure, and near-interface chemistry (Ni enrichment) in enabling dislocation nucleation/transmission.

***(i) <u>Deformation twinning in Ni</u>***

The appearance of deformation twins in the Ni-rich lamellae at higher strains suggests that dislocation slip alone is no longer sufficient to sustain continued plastic flow in the confined FCC phase. Although room temperature deformation of Ni is generally slip-dominated owing to its relatively high stacking fault energy (SFE) of ~130 mJ/m$^2$ [63], twinning can become competitive due to the high local stress in the confined ultrafine lamellar structure and, chemical effects due to substantial Cr in the Ni-rich FCC solid solution. The confined length scale does not directly promote twinning; instead, by reducing the dislocation mean free path and increasing the stress required for continued glide, it elevates the local driving force for the successive emission of twinning partials, thereby favoring deformation twinning at higher strains [64]. The effect of Cr is likely more direct. Cr retained in solid solution within the Ni-rich lamellae is expected to reduce the intrinsic SFE of the FCC Ni matrix, thereby promoting dissociation of perfect dislocations into Shockley partials. Recent first-principles calculations identifies Cr as an effective solute for lowering the SFE of Ni, while Cr-containing Ni alloys also show a broader spread of local SFEs due to local charge distortion [65, 66]. Such a reduction, and local heterogeneity, in fault energy would favor planar faults and the successive emission of partials on neighboring $\{1\ 1\ 1\}$ planes, thus making deformation twinning increasingly competitive to glide once the local stress becomes sufficiently high.

The SFE ($\gamma$) was further estimated directly from the faulted configuration in Fig. 7a by measuring the partial-dislocation separation (i.e., average length of stacking faults, $l_{sf}$) in regions representing the transition from perfect-dislocation-mediated plasticity to twinning, where faulted partial activity had just started to develop (zone B to zone A transition under nanoindentation in Figure 6b). Such regions were specifically selected to minimize the influence of extensive twin thickening and to better capture the local energetics governing partial-dislocation formation in the Ni-rich phase. The following model has been invoked to experimentally determine the SFE [67]:

$$\gamma = \frac{\mu b_p^2}{8\pi l_{sf}}\left(\frac{2-\upsilon}{1-\upsilon}\right)\left(1-\frac{2\upsilon\cos 2\theta}{2-\upsilon}\right) \quad (23)$$

Where $G$ represents shear modulus, $b_p$ is the magnitude of Burgers vectors of the Shockley partial dislocations, $\upsilon$ denotes Poisson's ratio, and $\theta$ is the angle between Burgers vector and dislocation line direction. As identified from *Equation (4)*, μ for the Ni-rich solid solution is $\approx 91\ GPa$, $b_p$ for a Shockley partial in Ni with Burgers vector $\frac{a}{6}\langle 1\ 1\ 2\rangle$ is $\approx 0.144\ nm$, $\upsilon$ for most metallic materials is $\approx 0.3$, and $\theta$ is $90°$ for an edge dislocation. Measured $l_{sf}$ from figure 7a is $\approx 26 \pm 7\ nm$. Therefore, the calculated SFE is in the range of $\approx 13 - 22\ mJ/m^2$. For comparison, the expected SFE was also estimated indirectly following the approach adopted by Dodaran *et al*. [68], using XRD-derived stacking-fault probability, i.e., a dimensionless measure of the probability of deformation faulting in the FCC stacking sequence (manifested as stacking-fault-density) reported by Deléhouzée *et al*. [69] and assuming an inverse relation between SFE ($\gamma$) and stacking fault density ($\alpha$) given as:

$$\gamma \propto \frac{1}{\alpha} \tag{24}$$

i.e., $$\gamma_{alloy} = \gamma_{Ni} \times \frac{\alpha_{Ni}}{\alpha_{alloy}} \tag{25}$$

As predicted by Deléhouzée *et al.*, $\alpha_{Ni}$ (for pure Ni) and $\alpha_{alloy}$ (~Ni–39Cr) are $2 \times 10^{-3}$ and $14 \times 10^{-3}$, respectively. $\gamma_{Ni}$ is frequently reported as $\approx 130\ mJ/m^2$. Thus, $\gamma_{alloy}$ (Ni–39Cr) is $\approx 18.5\ mJ/m^2$, which falls well within the limit estimated using *Equation (23)*. These SFE values should be regarded as effective local estimates rather than exact thermodynamic quantities. Taken together, the literature-based and TEM-based estimates indicate that Cr lowers the SFE of the Ni-rich phase.

The stacking fault energy (SFE) dependence on Cr concentration in Ni was further calculated via Density Functional Theory (DFT), as shown in Fig.7b. The set of slab model consists of 9 atomic layers and a 12 Å vacuum region along y-direction to eliminate periodic interactions, as shown in the inset of Fig.7b. Periodic boundary conditions were applied in both x and z directions. The *x*-, *y*-, *z*- axes lied along $[11\bar{2}]$, $[111]$, $[1\bar{1}0]$, respectively. In each slip atomic layer (adjacent to the short-dashed line), a Ni atom is substituted by a Cr atom. The Cr concentration is adjusted by varying the total number of atoms per layer. For example, a layer of 16 atoms corresponds to a Cr concentration of 6.25%, while a layer of 8 atoms yields 12.5%. The "fault" model was generated by shearing the upper crystal of the "perfect" model along the partial Burgers vector $^{1}/_{6}\langle 112\rangle$. The Intrinsic stacking fault energy (ISFE) was calculated based on the energy difference between the "perfect" and "faulted" models. For simplicity, only the configuration where the two Cr solute atoms are first-nearest neighbors (1$^{st}$-NN) was considered in perfect models. Upon shearing (faulting), the solute pair can be transformed into a 1$^{st}$-NN or a 2$^{nd}$-NN configuration. Considering the contribution of various solute-solute configurations to the SFE [70], we estimated the probability that a perfect model containing a 1$^{st}$-NN solute pair transforms into a faulted model with a 1$^{st}$-NN or 2$^{nd}$-NN relation to be 2/3 and 1/3, respectively. Correspondingly, the ISFE is calculated as:

$$\gamma_{\mathrm{ISF}} = \frac{2}{3}\gamma_{\mathrm{ISF}}^{(11)} + \frac{1}{3}\gamma_{\mathrm{ISF}}^{(12)} \tag{26}$$

Where $\gamma_{\mathrm{ISF}}^{(11)}$ and $\gamma_{\mathrm{ISF}}^{(12)}$ denote the ISFE associated with a 1$^{st}$-NN or a 2$^{nd}$-NN configuration after shearing from a 1$^{st}$-NN configuration. As shown in Fig. 7b, the calculated ISFE decreases monotonically with increasing Cr concentration, which is in excellent agreement with previously reported experimental and theoretical values [68, 69, 71]. The results support the conclusion that Cr promotes partial-dislocation activity and thereby facilitates deformation twinning in the Ni-rich phase.

Once activated, deformation twinning in the Ni-rich lamellae contributes directly to sustained hardening at large plastic strain by continuously increasing the density of barriers to dislocation motion (Figure 4d1 inset). As twins form and thicken, evidenced by the hierarchical twin structure, they progressively subdivide the FCC lamellae and reduce the effective dislocation mean free path, thereby increasing the stress required for continued glide [72]. The observed twin–dislocation reactions further indicate that twin boundaries act as effective sites for dislocation blocking and storage, while twin–twin interactions create highly constrained regions for subsequent plastic flow. Moreover, the formation of Lomer–Cottrell-like sessile locks at twin intersections introduces strong local obstacles that are difficult to bypass [73]. Together, these mechanisms suppress recovery, delay strain localization, and enable the Ni-rich phase to maintain substantial work hardening to large plastic strain. Notably, although significant hardening is retained during Stage III, the work-hardening-rate versus stress plot exhibits a pronounced change in slope beyond a critical stress (Figure 3a), indicating a transition in the dominant deformation response. This change is consistent with redistribution of load from the Ni-rich phase and the onset of plastic deformation in the Cr-rich BCC lamellae, whose origin and hardening contribution are discussed in the following section.

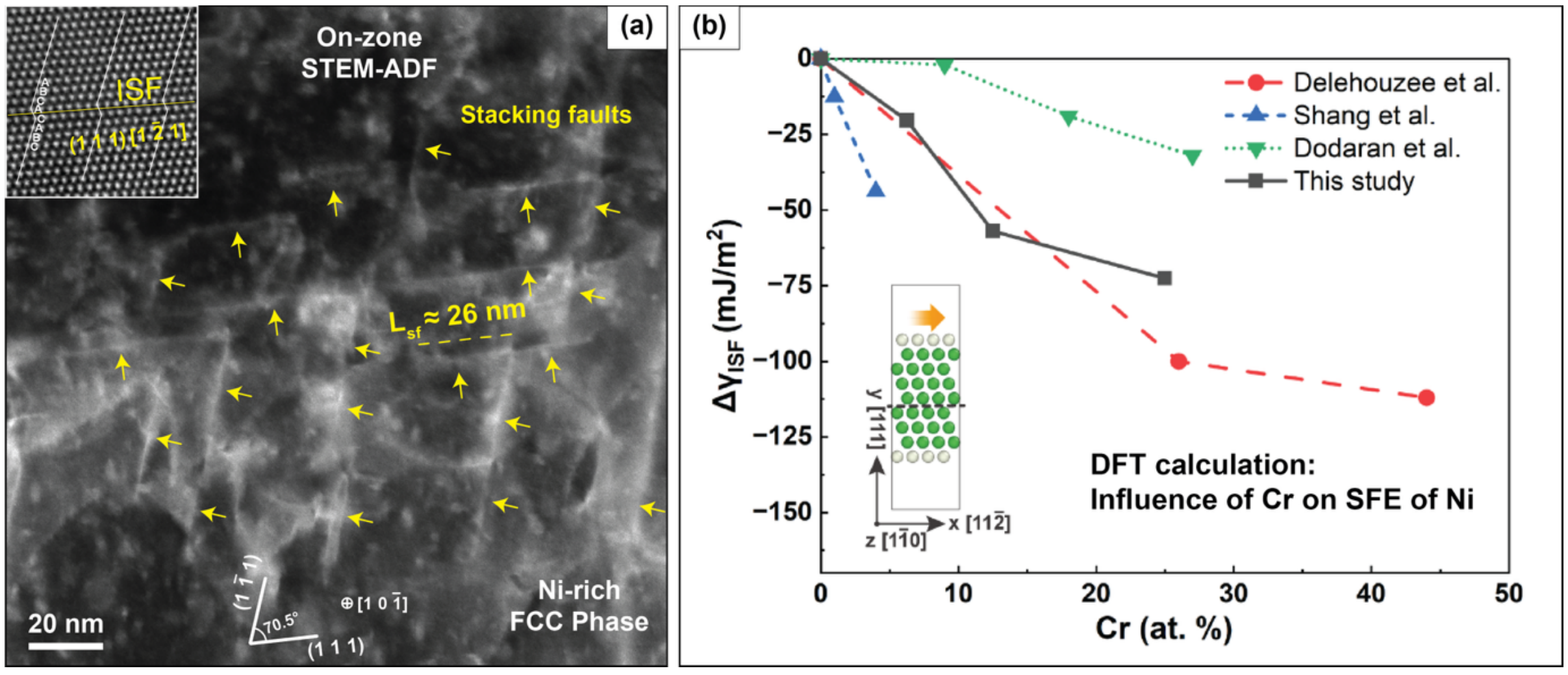

Figure 7: (a) On-zone STEM-ADF image of the faulted configuration in the Ni-rich lamellae, taken from the transition regime between perfect-dislocation-mediated plasticity and twinning; the partial-dislocation separation was used to estimate the effective local SFE. (b) DFT-calculated intrinsic SFE of FCC Ni as a function of Cr concentration, demonstrating a monotonic reduction in SFE with increasing Cr addition. The results support the role of Cr in facilitating partial-dislocation activity and deformation twinning in the Ni-rich phase.

### *(ii)* ***<u>Plasticity in Cr-rich BCC phase: Effects of interface orientation and chemistry</u>***

The next key question is how the Cr-rich BCC lamellae, which would otherwise be expected to exhibit limited plasticity at room temperature [13], become actively involved in deformation at higher strains. The first enabling factor is the ultrafine Cr lamellar length scale embedded within a plastically flowing Ni matrix. Once Ni yields, compatibility requires that strain be transferred across interfaces over a distance comparable to the lamellar spacing, so the Cr lamellae experience strong local stress concentrations and strain gradients even if their average macroscopic strain is smaller [74, 75]. In this microstructure, the near-interface region is therefore the natural site for dislocation activation in the hard phase. The GPA maps (Figure 5e–f) provide an independent indication that the Cr-rich phase does not remain purely elastic: the post-deformation strain fields are substantially higher than in the undeformed state and exhibit heterogeneous compressive/tensile components expected for constrained co-deformation in a lamellar composite.

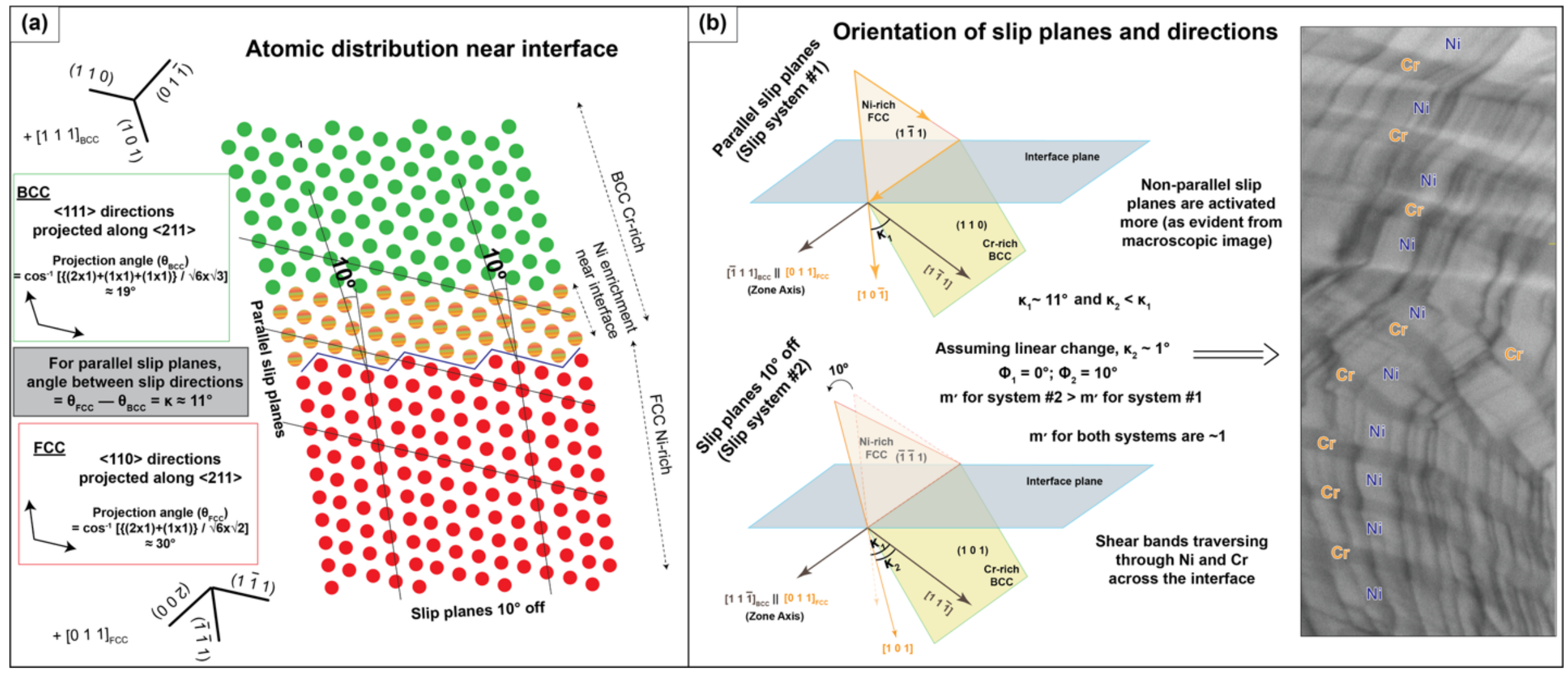

Figure 8: (a) Schematic representation of atomic distribution across the interface and the respective slip planes of FCC and BCC. One set of slip planes are parallel to each other while the other planes are 10° off. Based on the projection angles of slip directions, the angle between them ($\kappa_1$) for the parallel planes has been identified as ~11°. (b) 3D schematics of parallel slip planes ($\theta_1$ = 0°) (system #1) and 10° off slip planes ($\theta_2$ = 10°) (system #2). Assuming a linear rotation (valid for low angles), angle between slip directions for system #2 ($\kappa_2$) is calculated as ~1°. Since Luster–Morris parameter ($m' = \cos\phi \cos\kappa$) is close to 1 for both of the systems, slip transfer is feasible. The observed macroscopic shear-band trajectory is more consistent with activation of system #2, suggesting that this pathway is preferentially activated during co-deformation owing to slightly higher $m'$, although the small difference should not be interpreted as uniquely selecting one transfer mode.

The most important attribute for strain transfer to Cr is the FCC/BCC interface that satisfies the K–S OR (Figure 1f–g), which helps co-deformation primarily by making the FCC/BCC boundary an efficient slip-transfer and glide dislocation-generation pathway rather than a purely blocking barrier. HR-STEM shows a stepped, semi-coherent interface, implying a high density of crystallographically repeatable ledges/terraces and an ordered misfit structure. Under K–S, one dense FCC plane and direction pair is closely aligned with a dense BCC plane and direction pair, e.g., $\{1\ 1\ 1\}_{FCC} \parallel \{1\ 1\ 0\}_{BCC}$ and $\langle 0\ 1\ 1\rangle_{FCC} \parallel \langle 1\ 1\ 1\rangle_{BCC}$. However, the slip directions given here are along the zone axis, i.e., they are coming out of the plane of view, so they don't directly contribute to the slip transfer mechanism for the observable slip planes. Figure 8a illustrates that one set of slip planes are parallel to each other while the other planes are 10° off. Based on the projection angles of slip directions, the angle between them ($\kappa_1$) for the parallel planes has been identified as ~11°. 3D schematics (Figure 8b) of parallel slip planes ($\theta_1$ = 0°) (system #1) and 10° off slip planes ($\theta_2$ = 10°) (system #2) have been used for clearer visualization. Assuming a linear rotation (valid for low angles), angle between slip directions for system #2 ($\kappa_2$) is calculated as low as ~1°. Macroscopic shear bands help identify that the slip planes across the interface are predominantly activated over the other set of planes that are almost parallel to the interface.

The feasibility of slip transfer can be described using the Luster–Morris parameter:

$$m' = \cos\phi \cos\kappa \tag{27}$$

where $\phi$ and $\kappa$ are the angles between slip planes and between slip directions of the two phases. A larger $m'$ indicates better alignment of slip planes and directions and a smaller residual Burgers-vector penalty for transmission, which increases the geometric compatibility for slip transfer [76]. $m'$ is ~0.985 for system #2 and ~0.982 for system #1, both of them being close to its maximum value of 1, system #2 is likely favored, consistent with its slightly higher $m'$ and the observed macroscopic shear-band orientation. The Luster–Morris analysis, however, should not be interpreted as uniquely selecting one transfer pathway based only on the numerical difference in $m'$, rather, it indicates that the K–S FCC/BCC interface provides multiple geometrically favorable pathways for strain transfer. As $m'$ is close to 1, dislocations generated in the Ni-rich phase can more readily be absorbed at the interface and re-emitted into Cr, or trigger near-interface BCC dislocation nucleation when local stresses build up, because the interface needs less re-orientation of the shear. The stepped morphology further intensifies this effect: ledges act as sites that concentrate stress and provide discrete gates where an incoming FCC dislocation can be converted into an emitted BCC segment plus a residual interfacial component [77]. Mechanically, this interface-enabled activation is critical because it allows the Cr lamellae to share plastic strain with Ni once the incompatibility stresses become large, thereby reducing local tensile components in Cr that would otherwise promote cleavage and interlamellar cracking. In turn, the ability to continuously transfer/renew defects across a geometrically compatible K–S boundary helps sustain macroscopic shear-band continuity across lamellae and supports the observed stable co-deformation and sustained hardening at large strain.

Post-deformation interface characterization provides further evidence that the FCC/BCC boundary actively participates in plastic accommodation. In the as-built state, the interface follows a stepped semi-coherent K–S OR, with an average inclination of ~15° relative to the common/parallel planes. After deformation, the interface becomes macroscopically wavy (Figure 9a) and the local interface inclination increases to ~20°, while crystallographic analysis indicates a transition toward a Nishiyama–Wassermann (N–W) OR (Figure 9b) given as:

$$(1\ \bar{1}\ 1)_{FCC} \parallel (1\ 1\ 0)_{BCC};\ [0\ 1\ 1]_{FCC} \parallel [0\ 0\ 1]_{BCC}$$

The 5° change in interface inclination is consistent with deformation-induced interfacial shear/reorientation, as the K–S and N–W ORs in FCC/BCC systems are crystallographically close and can be connected through small rotations (~5°) associated with interfacial defect rearrangement [78]. Such reorientation indicates that accumulated interfacial shear is accommodated through rearrangement of interfacial steps/disconnections and residual dislocation content. These sheared interfacial regions provide favorable sites for both Shockley partial emission into the Ni-rich FCC phase (Figure 9c), producing interface-assisted twinning, and dislocation emission or transmission into the Cr-rich BCC phase (Figure 9d). Thus, the residual interfacial defect structure links the two key observations: twinning in Ni emanating from stepped interface regions and dislocation activity in Cr near the interface. Although the K–S interface is expected to be energetically favored in the undeformed state, the post-deformation N–W-like configuration should not be interpreted as a lower-energy equilibrium interface. Rather, it likely represents a deformation-stabilized interfacial state. During plastic co-deformation, accumulated interfacial shear and defect-transfer reactions can rotate/reconfigure the initially stepped K–S boundary toward an N–W-like alignment that better accommodates the imposed shear and reduces local incompatibility. The associated residual Burgers-vector content, interfacial steps/disconnections, and near-interface solute segregation can kinetically pin this reoriented structure after unloading, which has been illustrated in Figure 10. Thus, the N–W-like interface may be retained despite its higher static interfacial energy because it lowers the combined elastic/plastic mismatch energy of the deformed heterostructure.

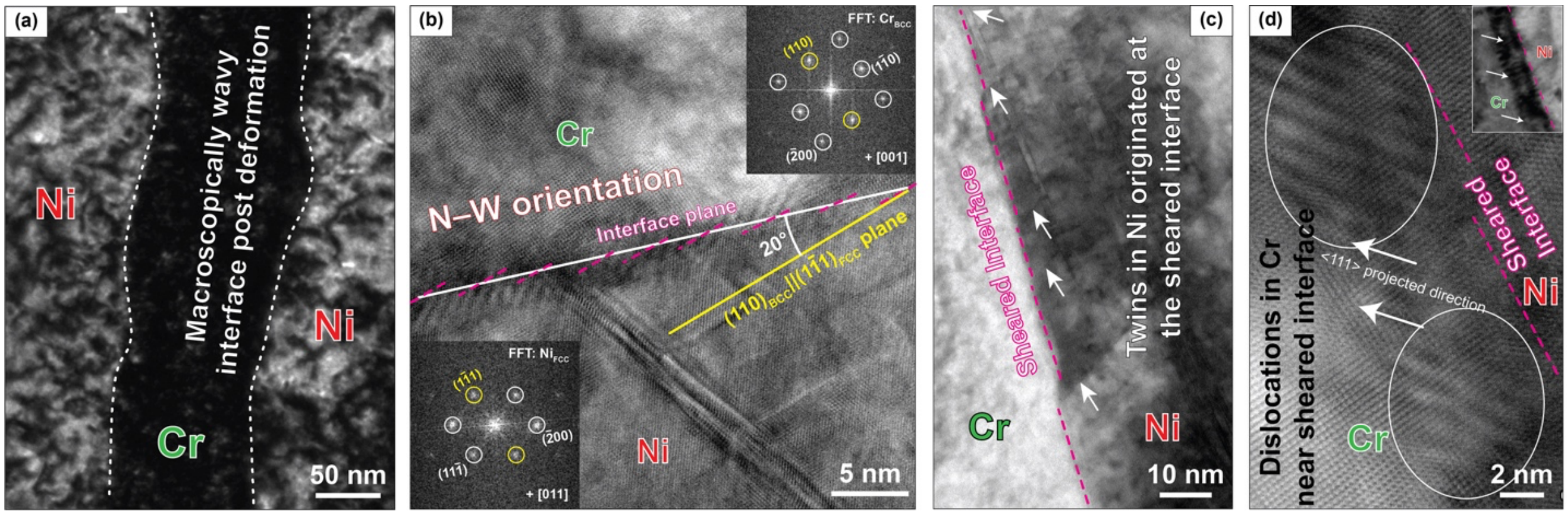

Figure 9. Post-deformation characterization of the FCC/BCC interface showing deformation-induced interfacial shear, local reorientation, and defect activity near the boundary. (a) Low-magnification STEM image showing a macroscopically wavy FCC/BCC interface after deformation, indicating accumulated interfacial shear accommodation. (b) HR-STEM image and corresponding FFT analysis showing that the post-deformation interface locally adopts a Nishiyama–Wassermann (N–W) OR, with the interface inclination increasing to ~20° relative to the common/parallel planes. (c) STEM image showing deformation twins in the Ni-rich FCC phase emanating from the sheared interface, suggesting interface-assisted Shockley partial emission and twin nucleation. (d) HR-STEM image showing deformation-induced dislocation activity in the Cr-rich BCC phase near the sheared FCC/BCC interface (inset: low magnification STEM image of the dislocations). Together, these observations indicate that the initially stepped K–S interface actively reconfigures during plastic co-deformation and serves as a site for residual defect storage, twinning in Ni-rich FCC, and dislocation activation in Cr-rich BCC.

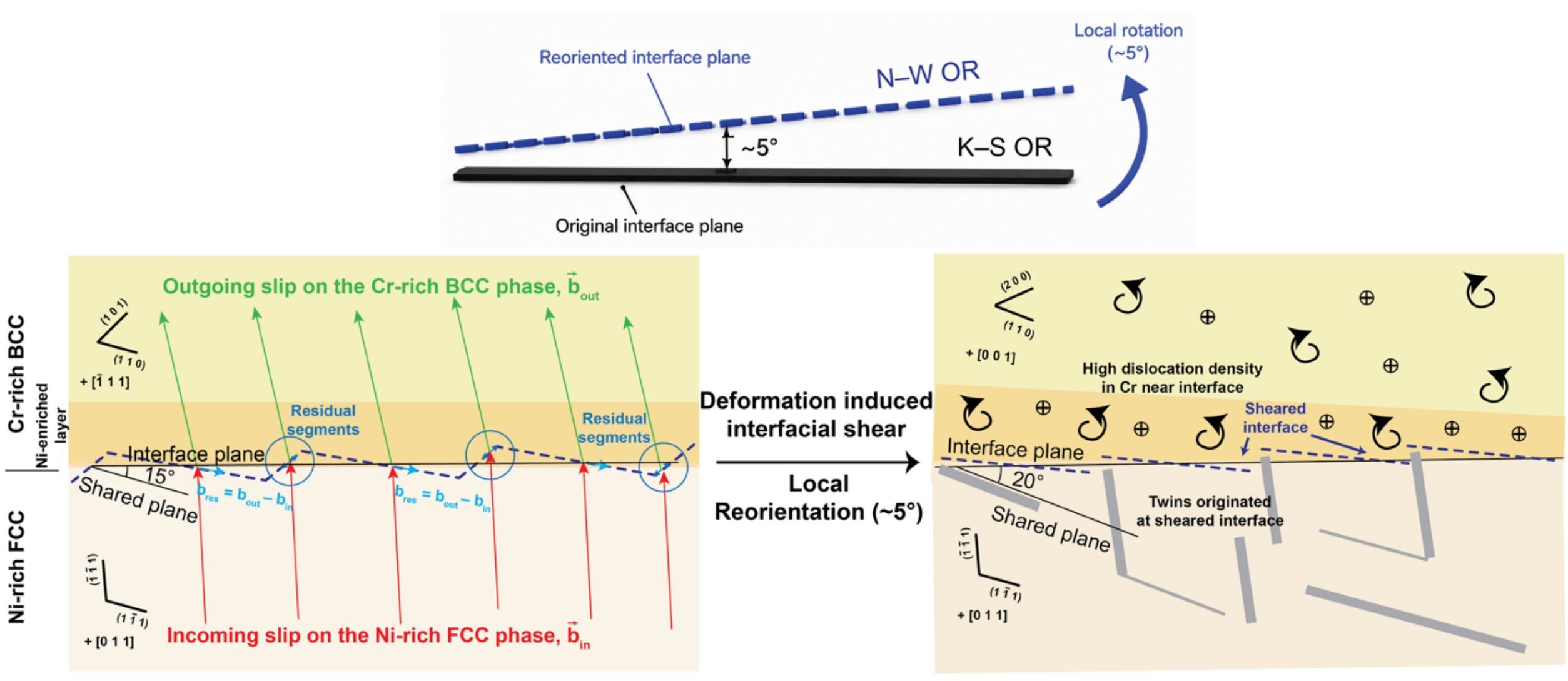

Figure 10. Schematic illustration of deformation-induced interfacial shear and local reorientation of the FCC/BCC interface from an initial K–S configuration toward an N–W-like configuration. In the as-built state, the stepped semi-coherent K–S interface contains periodic ledges/disconnections and an average inclination of ~15° relative to the common/parallel planes. During plastic deformation, incoming slip from the Ni-rich FCC phase interacts with the stepped interface, producing residual interfacial Burgers-vector content and promoting defect transfer across the FCC/BCC boundary. Accumulated interfacial shear locally reorients the boundary by ~5°, increasing the interface inclination to ~20° and producing an N–W-like post-deformation configuration. The sheared interface provides preferred sites for deformation twin nucleation into the Ni-rich FCC phase and dislocation activation in the Cr-rich BCC phase.

Another contributing factor is the near-interface Ni enrichment on the Cr side, which may influence the activation barrier for dislocation emission from the interface into Cr-rich BCC phase [79, 80, 81]. To directly assess whether the experimentally observed Ni enrichment on the Cr-rich side of the FCC/BCC interface can promote BCC plasticity, first-principles calculations were performed on Ni-modified BCC Cr configurations representative of the near-interface chemistry. The total energy of a dislocation on the $\{110\}\langle 111\rangle$ slip system is a combination of the long-range elastic strain energy ($E_{\text{elastic}}$) and the short-range core energy ($E_{\text{core}}$). The elastic component is governed by the effective shear modulus ($E \propto Gb^2$). Given the negligible difference in atomic radii between Cr and Ni, the variation in the Burgers vector $b$ is minor. Therefore, evaluating the directional shear modulus $G_{\{110\}\langle 111\rangle}$ allows for a direct assessment of how Ni segregation alters $E_{\text{elastic}}$, while the subsequent generalized stacking fault energy ($\gamma$-line) calculations address the $E_{\text{core}}$ component. To model the local chemical environments and inherent lattice distortions of the random alloy, five 72-atom Special Quasi-random Structure (SQS) [82] configurations were generated with Ni concentrations of 0, 5.56, 11.11, 16.67, and 22.22 at.%. The specific $G_{\{110\}\langle 111\rangle}$ values were extracted by fitting the energy-strain curves obtained from applied shear deformations. The calculated results, plotted in Figure 11a, reveal that $G_{\{110\}\langle 111\rangle}$ decreases as the Ni concentration rises. This trend indicates that Ni segregation effectively alleviates the long-range elastic strain energy of dislocations, establishing an initial energetic preference that favors enhanced plasticity from a continuum elasticity perspective.

The other vital component of dislocation energy, the core energy $E_{\mathrm{core}}$, is determined by the non-linear interatomic forces acting across the slip plane and is tied to the minimum misfit energy. According to the Peierls-Nabarro model, this misfit energy is mediated by the generalized stacking fault energy $\gamma$-line according to the following integral:

$$E_{\mathrm{misfit}} = \int_{-\infty}^{+\infty} \gamma\big(u(x)\big)\,\mathrm{d}x \tag{28}$$

Here, $u(x)$ the continuous misfit displacement at position $x$. To systematically investigate how Ni segregation affects the dislocation core, $\gamma$-line calculations were performed using an 8-layer supercell (9 atoms per layer, 72 atoms in total). A pure Cr reference model was compared to a targeted solute model where two Ni atoms occupy the first-nearest neighbor sites directly adjacent to the slip plane. Rather than describing this via a diluted global alloy concentration, this chemical substitution precisely reflects the highly concentrated local chemical environment at interface in EB-PBF Cr-Ni. The calculated $\gamma$-line profiles, plotted in Figure 11b, show that the stacking fault energy of pure Cr is higher than that of the Ni-modified model, confirming that Ni lowers the dislocation core energy barrier. Ultimately, the synchronized reductions in both the directional shear modulus $G_{\{110\}\langle 111\rangle}$ and the $\gamma$-line yield a definitive conclusion: Ni enrichment in the near-interface Cr-rich phase concurrently mitigates the long-range elastic strain field and the short-range atomic core barrier of dislocations. This synergetic energy relief substantially lowers the total dislocation self-energy, thereby lowering the energy barrier for dislocation emission on the Cr side.

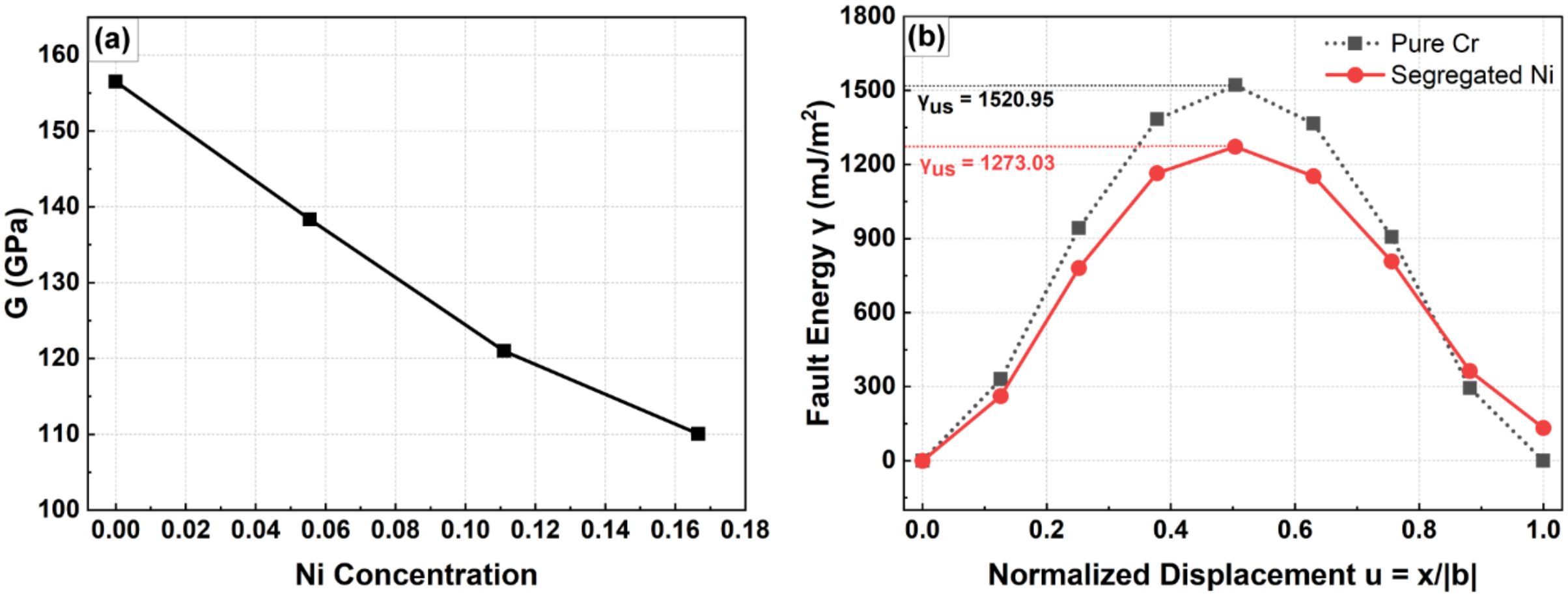

Figure 11 : (a) Directional shear modulus, G_({110}⟨111⟩ ), calculated as a function of Ni concentration in BCC Cr, showing a monotonic reduction with increasing Ni content. This reduction indicates that Ni enrichment lowers the long-range elastic strain energy associated with a/2 ⟨110⟩ dislocations. (b) DFT-calculated γ-line profiles for pure Cr and a Ni-segregated Cr model on the {110}⟨111⟩ slip system, showing a lower unstable fault energy for the Ni-modified configuration. Together, the reduction in directional shear modulus and γ-line barrier demonstrates that near-interface Ni enrichment lowers both the elastic and core-energy contributions to BCC dislocation nucleation, thereby promoting plasticity in the Cr-rich phase.

## 5. *Summary and conclusions*

This work demonstrates that an EPBF processed ultrafine Cr–Ni eutectic can achieve concurrent high strength and large plastic strain by enabling plastic co-deformation of Ni-rich FCC and Cr-rich BCC lamellae. The results show that the hard Cr-rich BCC phase, which would normally exhibit limited plasticity at room temperature, can be activated through nanoscale confinement, strain transfer, and interface-mediated defect mechanisms. The key conclusions are summarized below:

- EPBF processing produced an ultrafine lamellar Cr–Ni eutectic consisting of Cr-rich BCC and Ni-rich FCC phases with an average interlamellar spacing of ~450 nm. Atomic-resolution STEM revealed a stepped semi-coherent FCC/BCC interface obeying the K–S OR, with Ni enrichment confined to the near-interface region on the BCC side.
- *In situ* SEM nanomechanical testing demonstrated enhanced strength–plasticity response in both tension and compression. Compared with pure Ni and pure Cr reference specimens, the Cr–Ni eutectic exhibited substantially higher flow stress while sustaining large plastic strain without brittle failure of the Cr-rich phase.
- *Postmortem* TEM and strain-gradient nanoindentation revealed a sequential deformation pathway. Initial plasticity is dominated by dislocation glide and GND accumulation in the Ni-rich FCC lamellae near interfaces, followed by deformation twinning in the FCC phase and subsequent dislocation activity in the Cr-rich BCC lamellae at higher strain.
- The stepped FCC/BCC interface plays an active role in accommodating plastic incompatibility between the phases. Geometric compatibility across the K–S interface, deformation-induced interfacial shear and reorientation toward an N–W-like configuration, and residual interfacial defect content collectively promote slip transfer, interface-assisted twinning in Ni, and dislocation activation in Cr.
- Near-interface Ni enrichment on the Cr-rich side, synergistically with nanoscale lamellar confinement and the atomically stepped semi-coherent interface, helps reduce the barrier for dislocation emission in the hard phase from the interface, enabling stable plastic co-deformation below the monolithic brittle-to-ductile transition temperature of Cr.

## *Highlights*

- EB-PBF processing produced an ultrafine Cr–Ni eutectic with Cr-rich BCC and Ni-rich FCC lamellae separated by stepped semi-coherent interfaces.
- The ultrafine Cr–Ni eutectic exhibits a superior strength–plasticity combination compared with pure Cr and pure Ni reference specimens.
- Phase-resolved TEM reveals sequential deformation from Ni-rich FCC dislocation glide to FCC twinning and Cr-rich BCC dislocation activity.
- Interface-mediated strain transfer, local atomic reorientation, and near-interface Ni enrichment enable plastic co-deformation of the hard BCC phase.

## *Acknowledgements*

This work was funded by DOE, Office of Science, Office of Basic Energy Sciences with the grant number of DE-SC0016808. Arkajit Ghosh acknowledges the support of Rackham Predoctoral Fellowship by University of Michigan. Computational work at University of Nebraska-Lincoln was performed in the Nebraska Center for Materials and Nanoscience which is supported by the National Science Foundation under Award ECCS: 1542182 and the Nebraska Research Initiative. Experimental characterization was performed in the Michigan Center for Materials Characterization ($(MC)^2$) at the University of Michigan-Ann Arbor.

## *CRediT authorship contribution statement*

**Arkajit Ghosh:** Writing – original draft, Visualization, Methodology (testing and characterization), Investigation, Formal analysis, Data curation, Conceptualization. **Mustafa Tobah:** Methodology (specimen printing), Investigation. **Jianing Zhou:** Methodology (atomistic modeling), Validation, Formal Analysis, Software. **Jian Wang:** Writing – review & editing, Supervision, Project administration, Funding acquisition, Conceptualization. **Amit Misra:** Writing – review & editing, Supervision, Project administration, Funding acquisition, Conceptualization.